\title{MetaLP - IEEE Manuscript - Revised}

\documentclass[10pt,journal,compsoc]{IEEEtran}
\newcommand{\hLP}{\widehat{\LP}}

\ifCLASSOPTIONcompsoc
  \usepackage[nocompress]{cite}
\else
  \usepackage{cite}
\fi
\ifCLASSINFOpdf
  \usepackage[pdftex]{graphicx}
\else
\fi
\usepackage{amsmath}
\usepackage{url}

\usepackage{todonotes}
\usepackage{subcaption}
\usepackage{amsthm}
\usepackage{amssymb}
\usepackage{mathtools}
\usepackage{xargs}
\usepackage{scalerel}
\usepackage{hyperref}
\usepackage{stfloats,float,lscape}
\usepackage{multirow}
\usepackage[resetlabels]{multibib}
\usepackage{booktabs}
\usepackage{pifont}
\newcites{sec}{References}
\newtheorem{theorem}{Theorem}[section]

\theoremstyle{definition}
\newtheorem{definition}{Definition}[section]

\theoremstyle{remark}

\newcommand{\LP}{\operatorname{LP}}
\newcommand{\MLE}{\operatorname{MLE}}
\newcommand{\Fm}{F^{\rm{mid}}}
\newcommand{\Cor}{\operatorname{Cor}}
\newcommand{\beq}{\begin{equation}}
\newcommand{\eeq}{\end{equation}}
\def\bSig\mathbf{\Sigma}
\newcommand{\Ex}{\mathbb E}

\DeclarePairedDelimiterX{\expectarg}[1]{[}{]}{%
  \ifnum\currentgrouptype=16 \else\begingroup\fi
  \activatebar#1
  \ifnum\currentgrouptype=16 \else\endgroup\fi}
\newcommand{\innermid}{\nonscript\;\delimsize\vert\nonscript\;}
\newcommand{\activatebar}{%
  \begingroup\lccode`\~=`\|
  \lowercase{\endgroup\let~}\innermid 
  \mathcode`|=\string"8000
}
\newcommandx{\unsure}[2][1=]{\todo[linecolor=red,backgroundcolor=red!25,bordercolor=red,#1]{#2}}
\newcommandx{\change}[2][1=]{\todo[linecolor=blue,backgroundcolor=blue!25,bordercolor=blue,#1]{#2}}
\newcommandx{\info}[2][1=]{\todo[linecolor=OliveGreen,backgroundcolor=OliveGreen!25,bordercolor=OliveGreen,#1]{#2}}
\newcommandx{\improvement}[2][1=]{\todo[linecolor=Plum,backgroundcolor=Plum!25,bordercolor=Plum,#1]{#2}}
\newcommandx{\thiswillnotshow}[2][1=]{\todo[disable,#1]{#2}}
\newcommand{\cmark}{\ding{51}}%
\newcommand{\xmark}{\ding{55}}%

\begin{document}
%
\title{Nonparametric Distributed Learning Architecture for Big Data: Algorithm and Applications}
%
%
%
%

\author{Scott~Bruce, \textit{Student Member, IEEE,} Zeda Li, \textit{Student Member, IEEE,} Hsiang-Chieh~Yang, and~Subhadeep~Mukhopadhyay$^*$, \textit{Member, IEEE}
\IEEEcompsocitemizethanks{\IEEEcompsocthanksitem Temple University, Department
of Statistical Science \protect\\ Philadelphia,
PA, 19122.\protect\\
$^*$Corresponding Author E-mail: deep@temple.edu}

}

%
%

\markboth{IEEE TRANSACTIONS ON BIG DATA,~Vol.~**, No.~**, February~2018}%
{Shell \MakeLowercase{\textit{et al.}}: IEEE TRANSACTIONS ON BIG DATA}
%



\IEEEtitleabstractindextext{%
\begin{abstract}
Dramatic increases in the size and complexity of modern datasets have made traditional ``centralized'' statistical inference prohibitive.  In addition to computational challenges associated with big data learning, the presence of numerous data types (e.g. discrete, continuous, categorical, etc.) makes automation and scalability difficult.  A question of immediate concern is how to design a data-intensive statistical inference architecture without changing the basic statistical modeling principles developed for ``small'' data over the last century.  To address this problem, we present \texttt{MetaLP}, a flexible, distributed statistical modeling framework suitable for large-scale data analysis, where statistical inference meets big data computing.  This framework consists of three key components that work together to provide a holistic solution for big data learning: (i)  partitioning massive data into smaller datasets for \textit{parallel processing} and efficient computation, (ii) modern nonparametric learning based on a specially designed, orthonormal data transformation leading to \textit{mixed data algorithms}, and finally (iii) combining \textit{heterogeneous} ``local'' inferences from partitioned data using \textit{meta-analysis} techniques to arrive at the ``global'' inference for the original big data. We present an application of this general theory in the context of a nonparametric two-sample inference algorithm for Expedia personalized hotel recommendations based on $10$ million search result records.  


\end{abstract}

}

\maketitle

\IEEEdisplaynontitleabstractindextext

%
\IEEEpeerreviewmaketitle

\IEEEraisesectionheading{\section{Introduction}\label{sec:introduction}}

%
%
%
%

\IEEEPARstart{M}{\it {otivation}}. Expedia is a large online travel agency and has a strong interest in understanding how user, search, and hotel characteristics influence booking behavior. As a result, Expedia released a dataset \cite{KaggleData} containing $52$ variables of user and hotel characteristics (e.g.  search criteria, hotel features and pricing, user purchase history, competitor pricing, etc.) from $10$ million hotel search results collected over a window of the year $2013$. These factors will ultimately be used to optimize hotel search results and increase booking rates.  For this purpose, we develop a scalable, distributed algorithm that we refer to as \texttt{MetaLP}.  This learning algorithm can mine search data from millions of travelers, in a completely nonparametric manner, to find important features that best predict customers' likelihood to book a hotel.  This is an important large-scale machine learning problem.

{\it The Volume Problem}. This kind of ``tall" data structure, whose number of observations can run into the millions and billions, frequently arises in astronomy, marketing, neuroscience, e-commerce, and social networks. These massive datasets cannot be stored or analyzed by a single computer all-at-once using standard data analysis software.  This creates a major bottleneck for statistical modeling and inference. We seek to develop a framework that allows data scientists to systematically apply the tools and algorithms developed prior to the ``age of big data'' for massive data problems.

{\it The Variety Problem}. 
Another challenge is in developing a standard algorithm that can work across different data types, known as the mixed data problem \cite{ParzenMukhopadhyay2013LPM}. The Expedia dataset contains variables of different types (e.g. continuous, categorical, discrete, etc.), and  each requires a different statistical method for inference.  A few examples of traditional statistical measures for $(Y;X)$ data include: (1) Pearson's $\phi$-coefficient: $Y$ and $X$ both binary, (2) Wilcoxon statistic: $Y$ binary and $X$ continuous, (3) Kruskal-Wallis statistic: $Y$ discrete multinomial and $X$ continuous, and many more. Computational implementation of traditional statistical algorithms for large, mixed data thus become dauntingly complex as they require data type information to calculate the proper statistic. To streamline this process, we need to develop unified computing algorithms that yield appropriate statistical measures without demanding data type information from the user. To achieve this goal, we design a customized, discrete, orthonormal, polynomial-based transformation, the LP-Transformation \cite{MukhopadhyayParzen2014,Deep17LPMode}, suitable for arbitrary random variable $X$.  This transformation can be viewed as a nonparametric, data-adaptive generalization of Norbert Wiener's Hermite polynomial chaos-type representation \cite{Weiner1938}. This easy-to-implement LP-transformation based approach allows us to extend and integrate classical and modern statistical methods for nonparametric feature selection, thus providing the foundation to build automatic algorithms for large, complex datasets.

{\it The Scalability Problem}. Finally, the most crucial issue is to develop a scalable algorithm for large datasets, like the Expedia example. With the evolution of big data structures, new processing capabilities relying on distributed, parallel processing have been developed for efficient data manipulation and analysis. This paper presents a statistical inference framework for massive data that can fully exploit the power of parallel computing architecture and can be easily embedded into the MapReduce framework\cite{MapReduce04}. We design the statistical ``map'' function and ``reduce'' function for massive data variable selection by integrating many modern statistical concepts and ideas introduced in Sections \ref{sec:formulation} and \ref{sec:theory}. Doing so allows for faster processing of big datasets, while maintaining the ability to obtain accurate statistical inference without losing information. Another appealing aspect of our distributed statistical modeling strategy is that it is equally applicable for small and big data, thus providing a unified approach to modeling.

{\it Related Literature}. Several statistical distributed learning schemes for big data have been proposed in the literature. The divide and recombine (D\&R) approach \cite{GuhaCleveland2012,ClevelandHafen2014} to the analysis of large, complex data  provides a general statistical approach to analyzing big data in a way that is mostly considered embarrassingly parallel. In this setting, communication-efficient algorithms have been developed for various tasks such as assessing estimator quality \cite{Kleineretal2014}, statistical optimization \cite{Zhang2013}, and model aggregation \cite{Han2016}, based on bootstrap resampling and subsampling techniques.  Parallel algorithms have also been designed for large-scale parametric linear regression \cite{LinXi2011,ChenXie2014}.  These proposals address important challenges in analyzing large, complex data, but there are still significant hurdles to clear in developing a holistic framework for big data learning. Table \ref{tab:comparemethods} provides a comparison of the \texttt{MetaLP} learning framework with these proposals to better illustrate where this work fits into the existing distributed learning landscape.  

\begin{table*}[ht]
\begin{center}
\resizebox{14cm}{!}{
    \begin{tabular}{| l | c | c | c | c | c |}
    \hline
    Algorithm & Nonparametric & Inference & Modeling & Speed & Heterogeneity   \\ \hline
	MetaLP & \cmark & \cmark & \cmark & \cmark & \cmark \\
    BLB\cite{Kleineretal2014} & \cmark & \cmark & \xmark & \xmark & \xmark\\ 
    SAVGM\cite{Zhang2013} & \cmark & \cmark & \xmark & \xmark & \xmark \\ 
    KL-Weighting\cite{Han2016} & \xmark & \cmark & \cmark & \xmark & \xmark \\    
    AEE\cite{LinXi2011} & \xmark & \cmark & \cmark & \cmark & \xmark \\    
	Split-and-conquer\cite{ChenXie2014} & \xmark & \cmark & \cmark & \cmark & \xmark\\    
   \hline
    \end{tabular}
    }
\end{center}
\caption{Scope of \texttt{MetaLP} and other existing methods.}
\label{tab:comparemethods}
\end{table*}

The \texttt{MetaLP} framework broadens the existing scope of big data learning challenges that can be addressed in four important ways. First, the methods of \cite{LinXi2011,ChenXie2014} are based on parametric modeling assumptions. However, these assumptions often do not hold when analyzing large, complex data and present automation difficulties, as these assumptions are inherently data type dependent. On the other hand, the \texttt{MetaLP} framework is model--free in the sense that it is nonparametric and does not assume any specific model form. Thus, it is more applicable for big data analytics.  Second, while the communication-efficient algorithms \cite{Zhang2013,Han2016,Kleineretal2014}
are flexible to accommodate various statistics and model forms, they require specific instances from the user in order to conduct inference and modeling.  In contrast, our \texttt{MetaLP} framework relies on statistics based on the LP-transformation for nonparametric inference and modeling due to its favorable theoretical properties and its ability to solve the mixed data problem (see Section \ref{sec:theory}). Third,  \texttt{MetaLP} also enjoys a considerable reduction in computation time compared to other methods which rely on computationally intensive bootstrap resampling \cite{Kleineretal2014,Han2016} and subsampling \cite{Zhang2013} techniques in order to generate local inferences.  Lastly, characteristics across subpopulations may vary significantly, even under purely random data partitioning (see Section \ref{sec:BLBsim} and Supplementary Section \ref{sec:titanic}), which is known as heterogeneity \cite{HigginsThompson2002}.  Using meta-analysis principles, \texttt{MetaLP} assigns optimal weights to each local inference, which properly accounts for any potential heterogeneity.  These weights then determine the influence of each local inference on the final global inference.  This approach provides a crucial advantage over methods relying on equal weighting schemes where heterogeneity can spoil inference (see Sections \ref{sec:metacom} and \ref{sec:BLBsim}). In summary, this work provides the basis to develop a general and systematic massive data analysis framework that can simultaneously perform nonparametric statistical modeling and inference and can be adapted for a variety of learning problems.



{\it Organization}. Section \ref{sec:formulation} provides the basic statistical formulation and overview of the \texttt{MetaLP} algorithmic framework.  Section \ref{sec:theory} covers details of the individual elements of the distributed statistical learning framework, addressing the important issue of heterogeneity in big data along with a concrete nonparametric parallelizable variable selection algorithm.  Section \ref{sec:sim} evaluates the effectiveness of our proposed variable selection algorithm and compares it with other popular methods through simulation studies. Section \ref{sec:expedia} provides an in-depth analysis of the motivating Expedia dataset using the framework to conduct variable selection under different settings to determine which hotel and user characteristics influence booking behavior.   Section \ref{sec:conclusion} provides some concluding remarks and discusses the future direction of this work.  Supplementary materials are also available discussing two examples on how the \texttt{MetaLP} framework provides a new understanding and resolution for problems related to Simpson's Paradox and Stein's Paradox, the relevance of \texttt{MetaLP} for small-data, and the \texttt{R} scripts for MapReduce implementation.  

\section{Statistical Formulation of Big Data Analysis}\label{sec:formulation}

Our research is motivated by a real business problem of optimizing personalized web marketing for Expedia with the goal of improving customer experience and look-to-book ratios\footnote{The number of people who visit a travel agency web site compared to the number who make a purchase. This ratio measures the effectiveness of an agency in securing purchases.} by identifying key factors that affect consumer choices. This prototypical digital marketing case study allows us to address the following more general data modeling challenge, which finds its applicability in many areas of modern data-driven science, engineering, and business:

{\it How can we design nonparametric distributed algorithms that work on large amounts of data (that cannot be stored or processed by just one machine) to find the most important features that affect certain outcomes?}

\begin{figure} [ht] 
\centering
\includegraphics[height=\textheight,width=.35\textwidth,keepaspectratio,trim=2cm .5cm 2cm 1.5cm]{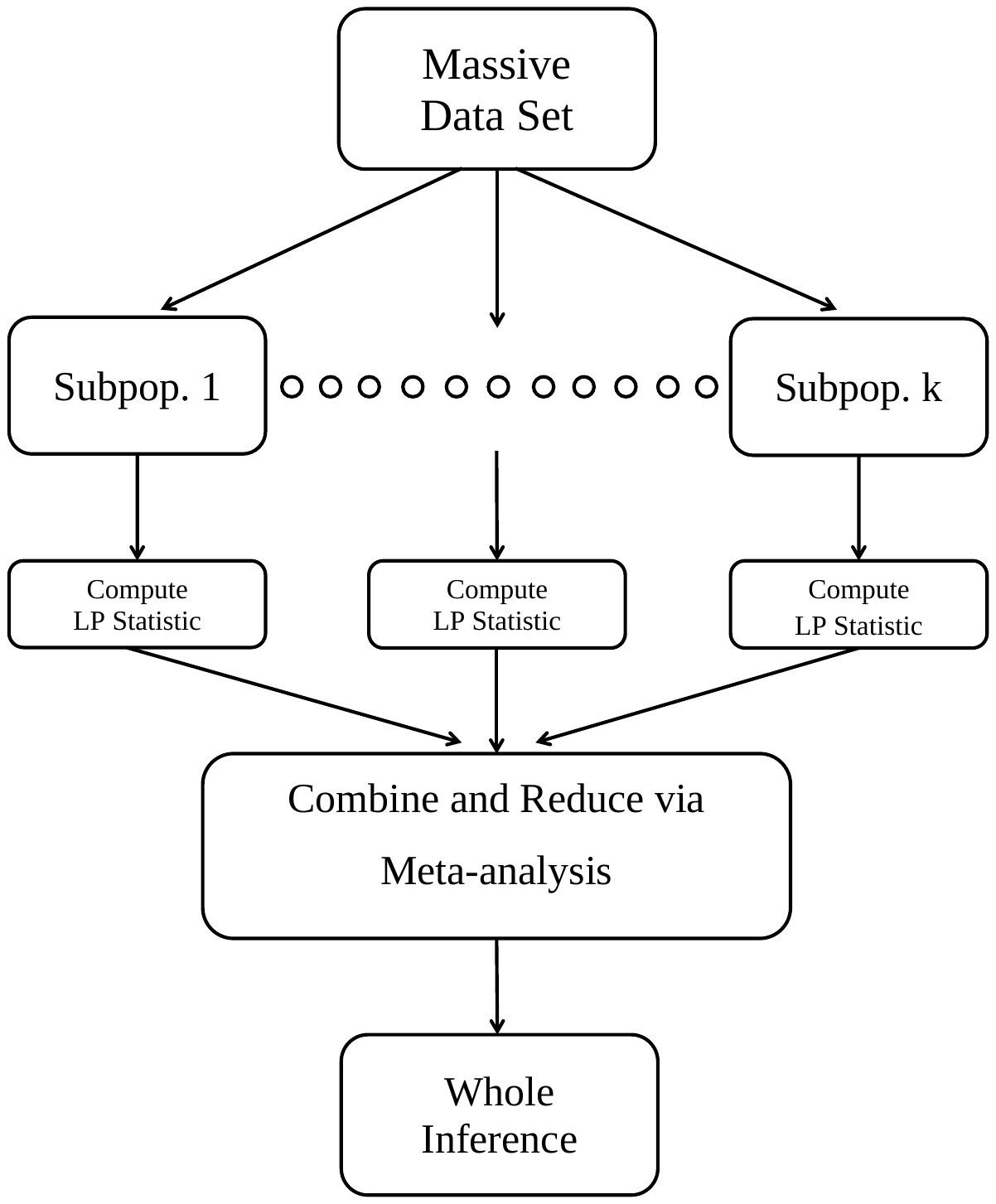}
 \caption{ \texttt{MetaLP} large-scale distributed statistical inference architecture.}
 \label {fig:meta_framework2}
\end{figure}

At first glance, this may look like a simple two-sample inference problem that can be solved by some trivial generalization of existing `small-data' statistical methods, but in reality, this is not the case. In this article we perform a thorough investigation of the theoretical and practical challenges present in big data analysis. We emphasize the role of statistics in big data analysis and provide an overview of the three main components of our statistical theory along with the modeling challenges they are designed to overcome. In what follows, we present the conceptual building blocks of \texttt{MetaLP}, a large-scale distributed learning tool that allows big data users to run statistical procedures on large amounts of data. Figure \ref{fig:meta_framework2} outlines the architecture.

\subsection{Partitioning Massive Datasets}
Dramatic increases in the size of datasets have created a major bottleneck for conducting statistical inference in a traditional, `centralized' manner, where we have access to the full data. The first, and quite natural, idea to tackle the volume problem is to divide the big data into several smaller datasets, similar to modern parallel computing database systems like Hadoop and Spark as illustrated in Figure \ref{fig:datatypes}. However, simply dividing the dataset does not allow data scientists to conquer the problem of big data analysis. There are many unsettled questions that we have to carefully address using proper statistical tools to arrive at an appropriate solution.

Users must select a data partitioning scheme to divide the original large data into several smaller parts and assign them to different nodes for processing.  The most common technique is random partitioning. However, users may choose other strategies, like spatial or temporal partitioning, in order to use the inherent structure of the data.  Also, the original massive dataset may already be partitioned by some natural grouping variable, in which case an algorithm that can accommodate pre-existing partitions is desirable. The number of partitions could also be defined by the user who may consider a wide range of cost metrics including the number of processors required, CPU time, job latency, memory utilization, and more.  

Another important, and often overlooked, consideration when choosing a partitioning scheme is that the characteristics of the subpopulations created may vary largely.  This is known as heterogeneity \cite{HigginsThompson2002} and is an unavoidable obstacle for divide-and-conquer style inference models.  Heterogeneity can certainly impact data-parallel inference, so we incorporate diagnostics to measure the severity of the problem (see Section \ref{sec:I2}) and data-adaptive regularization to adjust effect size estimates accordingly (see Section \ref{sec:htau}).  This allows users to detect the presence of heterogeneity in a given data partition and offers robustness to various data partitioning options in the estimation.

\begin{figure*}[!t]
\centering
\begin{subtable}{1\textwidth}
\centering
\caption*{Subpopulation 1}
\begin{tabular}{|c | c c c|}
\hline
\texttt{booking\_bool} & \texttt{promotion\_flag} & \texttt{srch\_length\_of\_stay} & \texttt{price\_usd} \\\hline
1 & 0 & 2 & 164.59 \\
0 & 1 & 7 & 284.48 \\
1 & 0 & 1 & 194.34 \\
\vdots & \vdots & \vdots & \vdots \\
0 & 1 & 3 & 371.27 \\
\hline
\end{tabular}
\end{subtable}


\[\bullet\]
\[\bullet\]
\[\bullet\]
\begin{subtable}{1\textwidth}
\centering
\caption*{Subpopulation $k$}
\begin{tabular}{|c | c c c|}
\hline
\texttt{booking\_bool} & \texttt{promotion\_flag} & \texttt{srch\_length\_of\_stay} & \texttt{price\_usd} \\\hline
1 & 1 & 1 & 125.65 \\
1 & 0 & 3 & 149.32 \\
0 & 1 & 1 & 224.46 \\
\vdots & \vdots & \vdots & \vdots \\
1 & 1 & 3 & 174.89 \\
\hline
\end{tabular}
\end{subtable}
\caption{Illustration of a partitioned data set with $k$ subpopulations and various data types.  Three variables in the {\it Expedia} dataset are shown.  The target variable $Y$, \texttt{booking\_bool}, indicates whether or not the hotel was booked. The three predictor variables shown are $X_1$ \texttt{promotion\_flag} (indicates if a sale price promotion was displayed), $X_2$ \texttt{srch\_length\_of\_stay} (search criterion for number of nights stayed), and $X_3$ \texttt{price\_usd} (displayed hotel price).}
\label{fig:datatypes}
\vspace{-.75em}
\end{figure*}

\subsection{LP Statistics for Mixed Data}\label{sec:LPMixedDataIntro}
Massive datasets typically contain a multitude of data types, and the Expedia dataset is no exception.  Figure \ref{fig:datatypes} shows three predictor variables with different data types in the Expedia dataset: \texttt{promotion\_flag} (binary), \texttt{srch\_length\_of\_stay} (discrete count), and \texttt{price\_usd} (continuous).  In order to construct appropriate statistical measures for identifying important variables, traditional algorithmic approaches demand two pieces of information: (1) values and (2) data type information for every variable.  This requirement produces considerable complications in computational implementation and creates serious roadblocks for building systematic and automatic algorithms for large, complex data. Thus, the question of immediate concern is:

{\it How can we develop a unified computing formula, with automatic built-in adjustments, that yields appropriate statistical measures, without requiring data type information from the user?}

To tackle this `data variety' or `mixed data' problem, we design a custom-constructed, discrete, orthonormal, polynomial-based transformation, called LP-Transformation.  This data transformation provides a generic and universal representation of any random variable, defined in Section \ref{sec:LP}. We use this transformation technique to represent the data in a new LP Hilbert space. This data-adaptive transformation will allow us to construct unified learning algorithms by compactly expressing them as inner products in the LP Hilbert space.

\subsection{Combining Information via Meta-Analysis}\label{sec:metacom}  Eventually, the goal of having a distributed inference procedure critically depends on the question:

{\it How to judiciously combine the `local' inferences executed in parallel by different servers to get the `global' inference for the original big data?}

To resolve this challenge, we make a novel connection with meta-analysis. Section \ref{sec:metaparallel} describes how we can use meta-analysis to parallelize the statistical inference process for massive datasets. Furthermore, we seek to provide a distribution estimator for the LP-statistics via a confidence distribution (CD) that contains information for virtually all types of statistical inference (e.g. estimation, hypothesis testing, confidence intervals, etc.). Section \ref{sec:CDmeta} discusses the use of CD-based meta-analysis, which plays a key role in integrating local inferences to construct a comprehensive answer for the original data. These new connections allow data scientists to fully utilize the parallel processing power of large-scale clusters for designing unified and efficient big data statistical inference algorithms.

To conclude, we have discussed the architectural overview  of \texttt{MetaLP}, which addresses the challenge of developing an inference framework for data-intensive applications without requiring modifications to the core statistical principles developed for `small' data. Next, we describe the theoretical underpinnings, algorithmic foundation, and implementation details of our data-parallel, large-scale \texttt{MetaLP} inference model.

\section{Elements of Distributed Statistical Learning}\label{sec:theory} 
In this section, we introduce the key concepts of our proposed method by connecting several classical and modern statistical ideas to develop a comprehensive inference framework. We highlight along the way how these new ideas and connections address the real challenges of big data analysis as noted in Section \ref{sec:formulation}.

\subsection{LP United Statistical Algorithm and Universal Representation}
\label{sec:LP}
To address the mixed data problem, we introduce a nonparametric statistical modeling framework based on an LP approach to data analysis \cite{MukhopadhyayParzen2014}. 

\textit{Data Transformation and LP Hilbert Functional Space Representation}.  Our approach relies on an alternative representation of the data in the LP Hilbert space, which will be defined shortly.  The new representation shows how each explanatory variable, regardless of data type, can be represented as a linear combination of {\it data-adaptive} orthogonal LP basis functions.  This data-driven transformation will allow us to construct unified learning algorithms in the LP Hilbert space. Many traditional and modern statistical measures developed for different data types can be compactly expressed as inner products in the LP Hilbert space. The following is the fundamental result for LP basis function representation.

\begin{theorem}[LP representation]
Random variable $X$ (discrete or continuous) with finite variance admits the following decomposition: $X - \Ex(X) = \sum_{j>0} \,T_j(X;X)\,  \Ex[X T_j(X;X)]$  with probability $1$.
\end{theorem}

$T_j(X;X)$, for $j=1,2,\ldots$, are score functions constructed by Gram Schmidt orthornormalization of the powers of $T_1(X;X)=\mathcal{Z}(\Fm(X;X))$. Where $\mathcal{Z}(X)=(X-\Ex[X])/\sigma(X)$, $\sigma^2(X) = \mbox{Var}(X)$, and the mid-distribution transformation of a random variable $X$ is defined as
\beq
\Fm(x;X)=F(x;X)-.5p(x;X)
\eeq
where $p(x;X)=\Pr[X=x], \,F(x;X)=\Pr[X \leq x]$. We construct the LP score functions on $0<u<1$  by letting $x=Q(u;X)$, where $Q(u;X)=\inf\{x:F(x) \geq u\}$ is the quantile function of the random variable $X$ and
\begin{align}
S_j(u;X)&=T_j(Q(u;X);X).
\end{align}

\begin{figure}[ht]
  \centering
 \includegraphics[scale=0.3,keepaspectratio,trim=1.5cm 0cm 1.5cm .5cm]{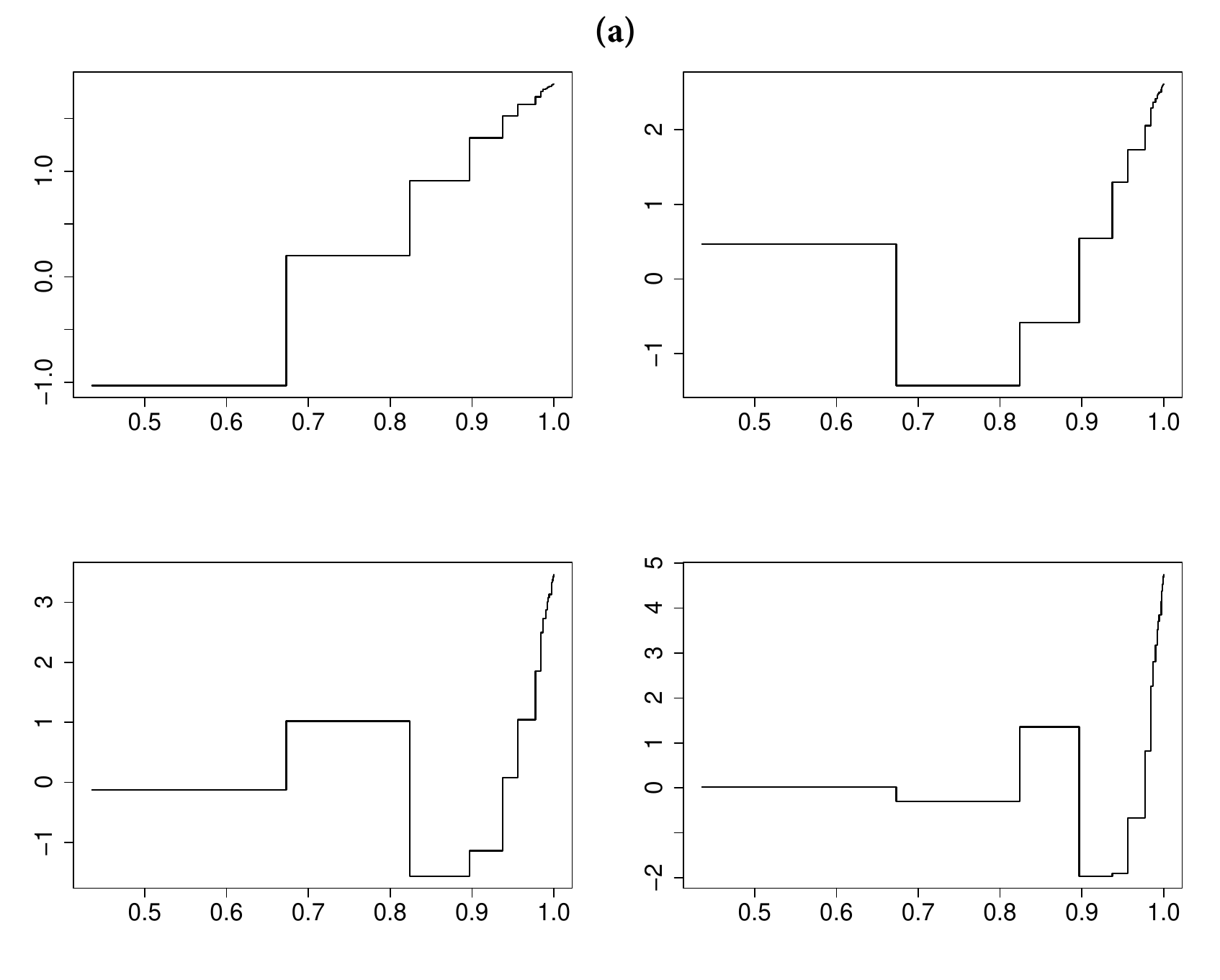}\\[1.5em]
 \includegraphics[scale=0.3,keepaspectratio,trim=1.5cm 0cm 1.5cm .5cm]{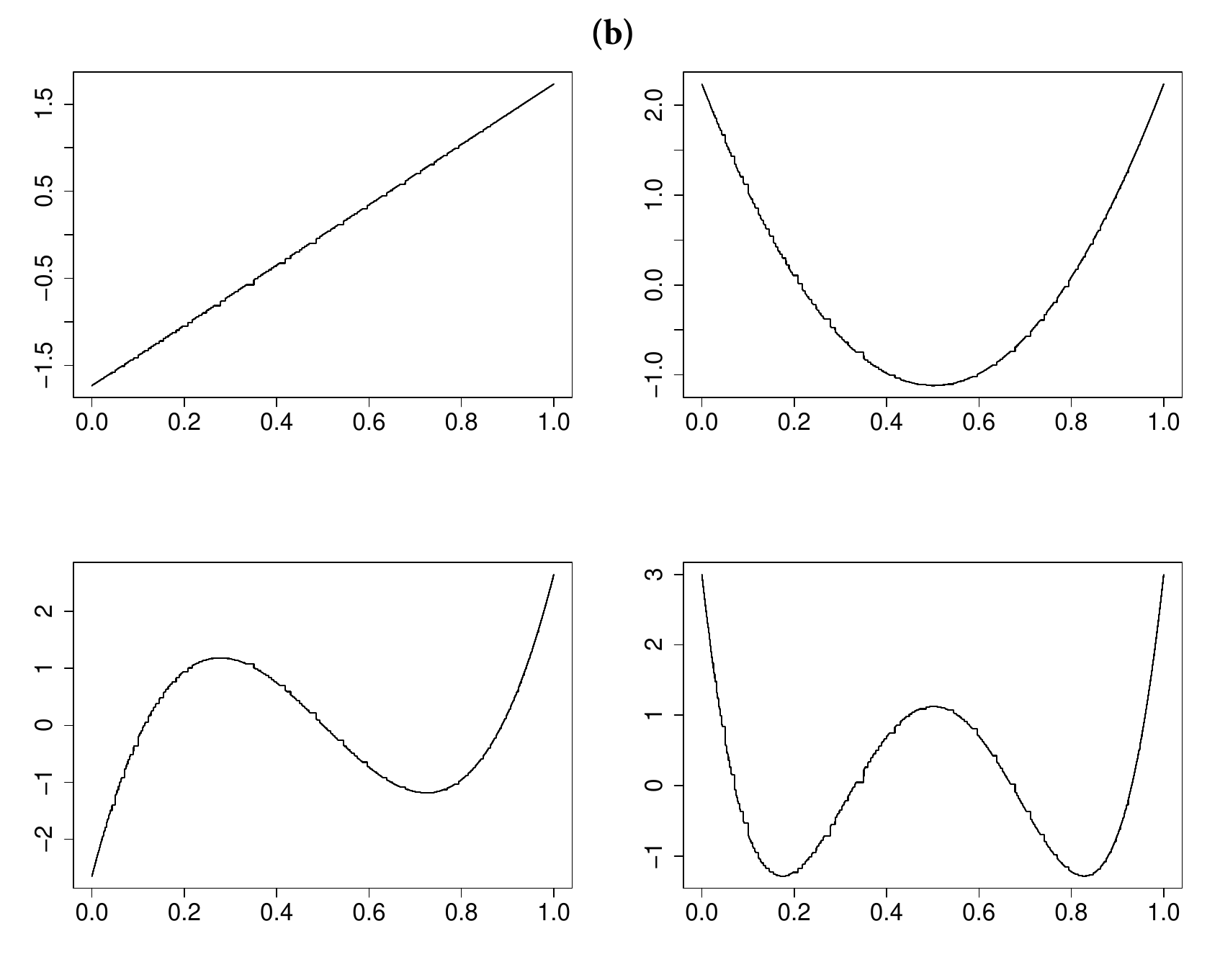}
 \caption{(a) Top $2 \times 2$ panel shows the shape of the first four LP orthonormal score functions for the variable \texttt{length\_of\_stay}, which is a discrete random variable taking values $0,\ldots,8$; (b) Bottom $2 \times 2$ panel shows the shape of the LP score functions for the continuous variable \texttt{price\_usd}.} \label{fig:score}
 \end{figure}

{\it Why is it called the LP-basis?} Note that our specially designed basis functions vary naturally according to data type unlike the fixed Fourier and wavelet bases as shown in Figure \ref{fig:score}. Note an interesting similarity of the shapes of LP score functions and shifted Legendre Polynomials for the {\it continuous} feature  \texttt{price\_usd}. In fact, as the number of distinct values of a random variable $A(X) \rightarrow \infty $ (moving from discrete to continuous data type), the shape converges to smooth Legendre Polynomials. To emphasize this universal limiting shape, we call it an {\bf L}egendre-{\bf P}olynomial-like ({\bf LP}) orthogonal basis. For any general $X$, LP-polynomials are piecewise-constant orthonormal functions over $[0,1]$, as shown in Figure \ref{fig:score}. This data-driven property makes the LP transformation uniquely advantageous in constructing a generic algorithmic framework to tackle the mixed data problem.  

\textit{Constructing Measures by LP Inner Product}. Define the two-sample LP statistic for variable selection of a \textit{mixed random} variable $X$ (either continuous or discrete) based on our specially designed score functions
\begin{align} \label{eq:LPm}
\LP[j;X,Y] &=\Ex[T_j(X;X)T_1(Y;Y)], \nonumber \\
&= \Cor[T_j(X;X),Y].
\end{align}
To prove Equation \eqref{eq:LPm}, which expresses our variable selection statistic as an LP-inner product measure, verify the following for $Y$ binary, 
\[\mathcal{Z}(y;Y)\,=\,T_1(y;Y)\,=\, \left\{ \begin{array}{rl}
 -\sqrt{\dfrac{p}{1-p}} &\mbox{for $y=0$} \\
  \sqrt{\dfrac{1-p}{p}} &\mbox{for $y=1$.}
       \end{array} \right.\]

\textit{LP statistic properties}. Using empirical process theory, we can show that the sample LP measures $\sqrt{n}\hLP[j;X,Y]$, asymptotically converge to i.i.d. standard normal  distributions \cite{MukhopadhyayParzen2014}.

As an example of the power of LP-unification, we describe $\hLP[1;X,Y]$ that systematically reproduces all the traditional linear statistical variable selection measures for different data types of $X$. Note that the nonparametric Wilcoxon method to test the equality of two distributions can equivalently be represented as $\Cor(\mathbb{I}\{Y=1\},\Fm(X;X))$, which leads to the following important alternative LP representation result.
\begin{theorem}
Two sample Wilcoxon Statistic $W$ can be computed as 
\beq W(X,Y)\,=\,\hLP[1;X,Y]. \eeq
\end{theorem}
Our computing formula for the Wilcoxon statistic using $\hLP[1;X,Y]$ offers automatic \textit{adjustments for data with ties}; hence, no further tuning is required. Furthermore, if we have $X$ and $Y$ both binary (i.e. data from the two variables can be represented in a $2 \times 2$ table), then we have 

\begin{eqnarray}
T_1(0;X)=-\sqrt{P_{2+}/P_{1+}}, & T_1(1;X)=\sqrt{P_{1+}/P_{2+}} \nonumber \\
T_1(0;Y)=-\sqrt{P_{+2}/P_{+1}}, & T_1(1;Y)=\sqrt{P_{+1}/P_{+2}},
\end{eqnarray}
where $P_{i+}=\sum_j P_{ij}$ and $P_{+j}=\sum_i P_{ij}$, and $P_{ij}$ denotes the entry for the $i$th row and $j$th column of the $2 \times 2$ probability table, and
\begin{align} \label{eq:phi}
\hLP[1;X,Y]&=\Ex[T_1(X;X) T_1(Y;Y)],\nonumber \\
&=\sum_{i=1}^2 \sum_{j=1}^2 P_{ij} T_1(i-1;X)\,T_1(j-1;Y), \nonumber \\
&=\big( P_{11}P_{22}-P_{12}P_{21}\big)/\big( P_{1+}P_{+1}P_{2+}P_{+2} \big)^{1/2}.
\end{align}
Following result summarizes the observation in \eqref{eq:phi}. 
\begin{theorem}
For a $2 \times 2$ contingency table with Pearson correlation $\phi$, we have,
\beq
\phi(X,Y)\,\,=\,\,\LP[1;X,Y].
\eeq
\end{theorem}

\textit{Beyond Linearity}. Higher order Wilcoxon statistics are LP statistics of higher order score functions, $T_j(X;X)$, which detect \textit{distributional} differences as in variability, skewness, or tail behavior for two different classes. The LP statistics $\LP[j;X,Y]$ for $j>1$ can capture how the distribution of a variable changes over classes and is applicable for mixed data types. 

To summarize, LP statistics allow data scientists to write a \textit{single} computing formula for any variable $X$, irrespective of its data type, with a \textit{common} metric and asymptotic characteristics. This leads to a huge practical benefit in designing a unified method for combining distributed 'local' inferences without requiring data type information for the variables.

\subsection{Meta-Analysis and Data-Parallelism}\label{sec:metaparallel} The objective of this section is to provide a new way of thinking about the problem: how to appropriately combine `local' inferences to derive reliable and robust conclusions for the original large dataset?  This turns out to be one of the most crucial, and heavily neglected, aspects of data-intensive modeling that decides the fate of big data inference. Here we introduce the required statistical framework that can answer the key question:  how to compute individual weights for each partitioned dataset? Our framework adopts the concept of meta-analysis to provide a general recipe for constructing such algorithms for large-scale parallel computing. This will allow us to develop statistical algorithms that can balance computational speed and statistical accuracy.

\textit{Brief Background on Meta-Analysis}. Meta-analysis \cite{HedgesOlkin1985} is a statistical technique by which information from independent studies is assimilated, which has  its origins in clinical settings. It was developed primarily to combat the problem of under-powered ``small data'' studies. A key benefit of this approach is the aggregation of information leading to improved statistical power as opposed to less precise inference derived from a single study.  A huge amount of literature exists on meta-analysis, including a careful review of recent developments \cite{SuttonHiggins2008}, which includes 281 references.

\textit{Relevance of Meta-analysis for big data inference?} Unlike the classical situation, we don't have statistical power issues for big data problems. However, we are unable to analyze the whole dataset all-at-once using a single machine in a classical inferential setup.  We apply meta-analysis from a completely different perspective and motivation, as a tool to facilitate distributed inference for massive datasets. This novel connection provides a statistically sound mechanism to combine ``local'' inferences by determining the optimal weighting strategy \cite{HedgesOlkin1985}.

We partition big data systematically into several subpopulations over a distributed database, estimate  parameters of interest in each subpopulation separately, and then combine results using meta-analysis as demonstrated in Figure \ref{fig:meta_framework2}. Thus, meta-analysis provides a way to pool information from subpopulations and produce a singular, powerful combined inference for the original large dataset. In some circumstances, the dataset may already be partitioned (e.g. each group could be an image or a large text document) and stored in different servers based on some reasonable partitioning scheme.  Our distributed statistical framework can work with these predefined groupings as well by combining them using the meta-analysis framework to arrive at the final combined inference.

We call this statistical framework, which utilizes both $\LP$ statistics and meta-analysis methodology, as \texttt{MetaLP}, and it consists of two parts: (i) the LP statistical map function or algorithm (that tackles the ``variety'' problem), and (ii) the meta-analysis methodology for merging the information from all subpopulations to get the final inference.

\subsection{Confidence Distribution and LP Statistic Representation}
\label{sec:CD}

The Confidence Distribution (CD) is a distribution estimator, rather than a point or interval estimator, for a particular parameter of interest. From the CD, all traditional forms of statistical estimation and inference (e.g. point estimation, confidence intervals, hypothesis testing) can be produced.  Moreover, CDs can be utilized within the meta-analysis framework, as we will show in the next section. More specifically, the CD is a sample-dependent distribution function on the parameter space that can represent confidence intervals of all levels for a parameter of interest. 

While the CD was first defined in \cite{SchwederHjort2002}, \cite{singh2005} extended the concept to asymptotic confidence distributions (aCDs). A comprehensive review of the concept can be found in \cite{XieSingh2013}.

\begin{definition}
Suppose $\Theta$ is the parameter space for an unknown parameter of
interest,  $\theta$, and $\omega$ is the sample space corresponding to
data $\mathbf{X}_n=\{X_1,X_2,\ldots,X_n\}^T$. Then a function
$H_n(\cdot)=H_n(\mathbf{X},\cdot)$ on $\omega \times \Theta \rightarrow
[0,1]$ is a confidence distribution (CD) if: (i) for each given
$\mathbf{X}_n\in \mathbf{\omega},H_n(\cdot)$ is a continuous cumulative
distribution function on $\Theta$; (ii) at the true parameter value
$\theta=\theta_0$, $H_n(\theta_0)=H_n(\mathbf{X},\theta_0)$, as a
function of the sample $\mathbf{X}_n$, following the uniform
distribution $U[0,1]$. The function $H_n(\cdot)$ is an asymptotic CD
(aCD) if the $U[0,1]$ requirement holds only asymptotically for  $n
\rightarrow \infty$ and the continuity requirement on $H_n(\cdot)$ can
be relaxed.
\end{definition}

The CD is a function of both a random sample and the parameter of interest. The additional requirement in (i) is that for each sample, the CD should be a distribution function on the parameter space. The $U[0,1]$ requirement in (ii) allows us to construct confidence intervals from a CD easily, meaning that $(H_n^{-1}(\alpha_1),H_n^{-1}(1-\alpha_2))$ is
a $100(1-\alpha_1-\alpha_2)\%$ confidence interval for the
parameter $\theta_0$ for any $\alpha_1>0$, $\alpha_2>0$, and
$\alpha_1+\alpha_2 <1$. 

Generally, the CD can easily be derived from the stochastic internal representation \cite{Parzen2013} of a random variable and a pivot, $\Psi(S,\theta)$.  The distribution of the pivot should not depend on the parameter, $\theta$, where $\theta$ is the parameter of interest and $S$ is a statistic derived from the data. Here, we derive the CD for the LP statistic.  Suppose $\widehat{\LP}[j;X,Y]$ is the estimated $j$th LP statistic for the predictor variable $X$ and binary response $Y$. The limiting asymptotic normality of the empirical LP statistic can be compactly represent as:
\beq
\LP[j;X,Y]
  \boldsymbol{\mathrel{\stretchto{\mid}{4ex}}} \widehat{\LP}[j;X,Y] =
 \widehat{\LP}[j;X,Y] + \frac{Z}{\sqrt{n}},
\eeq
which is the stochastic internal representation of the LP statistic, similar to the stochastic differential equations representation. Thus, we have the following form of the confidence distribution, which is the cumulative distribution function of $\mathcal{N}\big (\widehat{\LP}[j;X,Y], 1/n \big)$:
\beq
\label{eq:cd}
H_{\Phi}(\LP[j;X,Y])=\Phi\left( \sqrt{n}\left(\LP[j;X,Y]-\widehat{\LP}[j;X,Y]\right)\right).
\eeq
The above representation satisfies the conditions in the CD definition as $n \rightarrow \infty$ and therefore is the asymptotic CD of $\LP[j;X,Y]$.

\subsection{Confidence Distribution-based Meta-Analysis}\label{sec:CDmeta}
Using the theory presented in Section \ref{sec:CD}, we can estimate the CD for the $\LP$ statistics for each of the subpopulations, $H(\LP_\ell[j;X,Y]),$ and the corresponding point estimators, $\widehat{\LP}_\ell[j;X,Y]$, for $\ell=1,\ldots,k$.  The next step of our \texttt{MetaLP} algorithm is to judiciously combine information contained in the CDs for all subpopulations to arrive at the combined CD, $H^{(c)}(\LP[j;X,Y])$, based on the whole dataset for each specific variable $X$.  The framework relies on a confidence distribution-based approach to meta-analysis \cite{XieSinghStrawderman2011}.  The combining function for CDs across $k$ different studies can be expressed as:
\begin{align}
\label{eq:combinedcd}
&H^{(c)}(\LP[j;X,Y]) \nonumber\\
&=G_c \{g_c(H(\LP_1[j;X,Y]),\ldots,H(\LP_k[j;X,Y]))\}. 
\end{align}
The function $G_c$ is determined by the monotonic $g_c$ function: $G_c(t)=P(g_c(U_1,\ldots,U_k)\le t)$, where $U_1,\ldots,U_k$ are independent $U[0,1]$ random variables. A popular and useful choice for $g_c$ is
\beq \label{eq:com2}
g_c(u_1,\ldots,u_k)=\alpha_1F_0^{-1}(u_1)+\ldots+\alpha_kF_0^{-1}(u_k),
\eeq

where $F_0(\cdot)$ is a given cumulative distribution
function and $\alpha_\ell \ge 0$ , with at least one $\alpha_\ell \ne 0$, are generic weights. $F_0(\cdot)$ could be any distribution function, which highlights the flexibility of the proposed framework.  Hence, the following theorem introduces a reasonable proposed form of the combined aCD for $\LP[j;X,Y]$.

\begin{theorem}\label{thm:fixed}
Setting $F_0^{-1}(t) = \Phi^{-1}(t)$ and $\alpha_l = \sqrt{n_\ell}$, where $n_\ell$ is the size of subpopulation $\ell=1,\ldots,k$, the following combined aCD for $\LP[j;X,Y])$ follows:
\begin{align} &H^{(c)}(\LP[j;X,Y]) \nonumber \\ 
&= \Phi \left[ \left( \sum\limits_{\ell=1}^k n_\ell \right)^{1/2} \left(\LP[j;X,Y] - \widehat{\LP}^{(c)}[j;X,Y]\right) \right] \quad 
\end{align} 
with
\beq 
\quad \widehat{\LP}^{(c)}[j;X,Y] = \frac{ \sum_{\ell=1}^{k} n_\ell \widehat{\LP}_\ell[j;X,Y]}
{ \sum_{\ell=1}^{k} n_\ell}\eeq
where $\widehat{\LP}^{(c)}[j;X,Y]$ and $\left(\sum_{\ell=1}^k n_\ell\right)^{-1}$ are the mean and variance respectively of the combined aCD for $\LP[j;X,Y]$.
\end{theorem}

To prove this theorem, verify that replacing $H(\LP_\ell(j;X,Y))$ by \eqref{eq:cd} in Equation \eqref{eq:combinedcd} along with the choice of combining function given in \eqref{eq:com2}, where $F_0^{-1}(t) = \Phi^{-1}(t)$ and $\alpha_\ell = \sqrt{n_\ell}$, we have 
\begin{align*} 
&H^{(c)}(\LP[j;X,Y]) \nonumber \\
&= \Phi \left[ \frac{1}{\sqrt {\sum_{\ell=1}^{k}  n_\ell}} \sum\limits_{\ell=1}^k \sqrt{n_\ell} \frac{\LP[j;X,Y] - \widehat{\LP}_\ell[j;X,Y]}{1/\sqrt{n_\ell}}\right].
\end{align*}

\subsection{Diagnostic of Heterogeneity}
\label{sec:I2}
Heterogeneity is a common issue with divide, combine, and conquer approaches to big data analysis and is caused by different characteristics across subpopulations. This issue is often ignored and can easily spoil the big data discovery process by producing very different statistical estimates, which may not faithfully reflect the original parent dataset.  Therefore, we diagnose and quantify the degree to which each variable suffers from heterogeneous subpopulation groupings using the $I^2$ statistic \cite{HigginsThompson2002}. Define Cochran's Q statistic:
\beq \label{eq:Q}
Q\,=\, \sum_{\ell=1}^k \alpha_{\ell}\, \big(\widehat{\LP}_\ell [j;X,Y]-\widehat{\LP}^{(c)}[j;X,Y]\big)^2,
\eeq
where $\widehat{\LP}_\ell [j;X,Y]$ is the estimated $\LP$-statistic from subpopulation $\ell$, $\alpha_{\ell}$ is the weight for subpopulation $\ell$ as defined in Theorem \ref{thm:fixed}, and $\widehat{\LP}^{(c)}[j;X,Y]$ is the combined meta-analysis estimator. Compute the $I^2$ statistic by
\begin{equation}  \label{eq:I2}
I^2 = \left\{
  \begin{array}{l l}
    \frac{Q-(k-1)}{Q} \times 100 \% & \quad \text{if $Q>(k-1)$;}\\
    0 & \quad \text{if $Q \leq (k-1)$;}
  \end{array} \right.
\end{equation}
where $k$ is the number of subpopulations. As a general rule of thumb, $0\% \leq I^2 \leq 40 \%$ indicates heterogeneity among subpopulations is not severe.

\subsection {\texorpdfstring{$\tau^2$}{Lg} Regularization to Tackle Heterogeneity in Big Data} \label{sec:htau}

Variations among the subpopulations impact $\LP$ statistic estimates, which are not properly accounted for in the Theorem \ref{thm:fixed} model specification. This is especially severe for big data analysis, as it is very likely that a substantial number of variables may be affected by heterogeneity across subpopulations.  To better account for the heterogeneity in our distributed statistical inference framework, following \cite{HedgesOlkin1985}, we introduce an additional parameter, $\tau^2$, to account for uncertainty due to heterogeneity across subpopulations. This results in a hierarchical model structure:
\begin{equation}
\label{eq:r1}
\widehat{\LP}_\ell[j;X,Y]\; \boldsymbol{\mathrel{\stretchto{\mid}{4ex}}} 
 \;\LP_\ell[j;X,Y],s_i ~\stackrel{\mathrm{iid}}{\sim}~ N(\LP_\ell[j;X,Y],s_i^2), \eeq 
\beq \label{eq:r2}
\LP_\ell[j;X,Y]\; \boldsymbol{\mathrel{\stretchto{\mid}{4ex}}} 
 \; \LP[j;X,Y],\tau ~\stackrel{\mathrm{iid}}{\sim}~ N(\LP[j;X,Y],\tau^2),
\end{equation}
where $\ell=1,\ldots,k$. This model describes two sources of variability of the LP statistic: variation between different subpopulations, and sampling variability within each subpopulation. Note that when $\tau=0$, all the subpopulation LP effect size estimates, $\widehat{\LP}_\ell[j;X,Y]$, come from a \textit{single}, homogeneous distribution. Thus, when $I^2$ indicates the presence of ``excess''  variability among $\{\widehat{\LP}_1,\ldots, \widehat{\LP}_k\}$, beyond random fluctuation alone, it is important to introduce the second layer in \eqref{eq:r2} to account for that heterogeneity. See Sections \ref{sec:BLBsim} and \ref{sec:heter} for more discussion on this topic. 

Under the new model specification, the CD of the $\LP$ statistic for the $\ell$-th
group is $H(\LP_\ell[j;X,Y])=\Phi((\LP[j;X,Y]-\widehat{\LP}_\ell[j;X,Y])/(\tau^2 +s_\ell^2)^{1/2})$ where $s_\ell = 1/\sqrt{n_\ell}$. The following theorem provides the form of the combined aCD under this specification.

\begin{theorem} \label{thm:taulp}
Setting $F_0^{-1}(t) = \Phi^{-1}(t)$ and $\alpha_\ell = 1/\sqrt{(\tau^2+(1/n_\ell))}$, where $n_\ell$ is the size of subpopulation $\ell=1,\ldots,k$, the following combined aCD for $\LP[j;X,Y]$ follows:
\begin{multline*} 
\label{eq:CDcombine}
H^{(c)}(\LP[j;X,Y]) = \\
\Phi \left[ \left( \sum\limits_{\ell=1}^k \frac{1}{\tau^2+(1/n_\ell)} \right)^{1/2}
(\LP[j;X,Y] - \widehat{\LP}^{(c)}[j;X,Y]) \right] \;  
\end{multline*}  
with
\beq \label{eq:taulp}
\widehat{\LP}^{(c)}[j;X,Y]) = \frac { \sum_{\ell=1}^{k} (\tau^2+(1/n_\ell))^{-1} \widehat{\LP}_\ell[j;X,Y])}
{ \sum_{\ell=1}^{k} (\tau^2+(1/n_\ell))^{-1}}
\eeq
where $\widehat{\LP}^{(c)}[j;X,Y])$ and $(\sum_{\ell=1}^{k} 1/(\tau^2+(1/n_\ell)))^{-1}$ are the mean and variance respectively of the combined aCD for $\LP[j;X,Y]$.
\end{theorem}

\begin{table*}[t]
\begin{center}
\begin{tabular}{cc|c|c|c|c|c|c|c|}

\cline{1-8}
\multicolumn{1}{ |c| }{Observations}  &
\multicolumn{1}{ |c| }{Partition Parameter} & 
\multicolumn{1}{ |c| }{Mean Absolute} & 
\multicolumn{2}{ |c| }{Accuracy} & 
\multicolumn{2}{ |c| }{Run Time (seconds)} & 

\multicolumn{1}{ |c| }{Speed}\\ 

\cline{4-7}
\multicolumn{1}{ |c| }{$n$}  & 
\multicolumn{1}{ |c| }{$\gamma$} &
\multicolumn{1}{ |c| }{Error $(\times 10^5)$} & 
\multicolumn{1}{ |c| }{Full Data} & 
\multicolumn{1}{ |c| }{\texttt{MetaLP}} & 
\multicolumn{1}{ |c| }{Full Data} & 
\multicolumn{1}{ |c| }{\texttt{MetaLP}} & 
\multicolumn{1}{ |c| }{Increase} \\ 

\cline{1-8}
\multicolumn{1}{ |c  }{} &
\multicolumn{1}{ |c| }{0.3} & 79.87 &  & 1  &  & 0.09  & 10.9    \\ 
\multicolumn{1}{ |c  }{\multirow{1}{*}{5,000}}                        &
\multicolumn{1}{ |c| }{0.4} & 119.69& 1 & 1 & 0.98 & 0.08  & 12.3     \\ 
\multicolumn{1}{ |c  }{}                        &
\multicolumn{1}{ |c| }{0.5} & 189.43&  & 1 &  & 0.05 & 19.6  \\ 

\cline{1-8}
\multicolumn{1}{ |c  }{ } &
\multicolumn{1}{ |c| }{0.3} & 10.66 &  & 1  &  & 0.38   & 20.9    \\ 
\multicolumn{1}{ |c  }{\multirow{1}{*}{50,000}}                        &
\multicolumn{1}{ |c| }{0.4} & 19.69 & 1 & 1 & 7.95 & 0.20   & 39.8   \\ 
\multicolumn{1}{ |c  }{}                        &
\multicolumn{1}{ |c| }{0.5} & 34.42  &  & 1  &  & 0.13  & 61.2   \\ 

\cline{1-8}
\multicolumn{1}{ |c  }{ } &
\multicolumn{1}{ |c| }{0.3} & 1.56&  & 1&  & 2.02   & 41.4   \\ 
\multicolumn{1}{ |c  }{\multirow{1}{*}{500,000}}                        &
\multicolumn{1}{ |c| }{0.4} & 3.20 & 1 & 1 & 83.66 & 1.01  & 82.8     \\ 
\multicolumn{1}{ |c  }{}                        &
\multicolumn{1}{ |c| }{0.5} & 6.20&  & 1&  & 0.67   & 124.9     \\ 

\cline{1-8}
\multicolumn{1}{ |c  }{ } &
\multicolumn{1}{ |c| }{0.3}& 0.89 &  & 1 &  & 3.36   & 53.2  \\ 
\multicolumn{1}{ |c  }{\multirow{1}{*}{1,000,000}}                        &
\multicolumn{1}{ |c| }{0.4}  & 1.86 & 1 & 1 & 178.65 & 1.59 & 112.4  \\ 
\multicolumn{1}{ |c  }{}                        &
\multicolumn{1}{ |c| }{0.5} & 3.19 &  & 1 &  & 1.17  & 152.7    \\ 

\cline{1-8}
\end{tabular}
\end{center}
\caption{Comparison of estimation and run times for full data and \texttt{MetaLP} LP statistic estimation. Mean absolute error is reported for the Meta LP estimates of the full data LP statistics across all variables.  Accuracy is defined as the proportion of replications correctly selecting the true model variables, $\{X_1, X_2, X_3\}$.  Run times are reported along with the speed increase using the distributed \texttt{MetaLP} approach.
}  
\label{tab:tpfp}
\end{table*}
The proof is similar to that for Theorem \ref{thm:fixed}. The DerSimonian and Laird \cite{DerSimonianLaird1986} and restricted maximum likelihood estimators of the data-adaptive heterogeneity regularization parameter $\tau^2$ are provided in Supplementary Section \ref{sec:tau2estimator}.

\section{Simulation Studies} 
\label{sec:sim}

In our simulation studies, we investigate the performance of our \texttt{MetaLP} approach compared with the oracle full data $\LP$ estimates, as well as with existing methods.  We evaluate the methods from four perspectives: 1) accuracy in estimating the oracle full data $\LP$ statistics, 2) ability to correctly classify important variables and noise variables, 3) computational efficiency in terms of run time, and 4) performance under the influence of heterogeneity. The dataset we considered has the form $(\mathbf{X}_i, Y_i) \sim P$, i.i.d, for $i=1,2,...,n$, where $\mathbf{X}_i \in \mathbb{R}^p$ and $Y_i \in (0,1)$. We generate dataset from the model $Y_i \sim \text{Bernoulli}(P(\beta_1 X_{1i}^2 + \mathbf{X}^T_{(-1)i} \boldsymbol{\beta}_{-1} ))$, where $P(u) = \exp(u)/(1+\exp{(u)})$. $\mathbf{X}_{-1}$ and $\boldsymbol{\beta}_{-1}$ mean all $X$'s except $X_1$ and all $\beta$'s except $\beta_1$. We set $\boldsymbol{\beta} = (\beta_1, \beta_2,..., \beta_p)=(2, -1.5, 3, 0,...,0)^T$ to be a $p$-dimensional coefficient vector, where $p=50$, and then generate three important variables, $X_{1i}$, $X_{2i}$, and $X_{3i}$, from StudentT(30), Binomial($n=15,p=0.1$), and Bernoulli($p=0.2$) respectively. The remaining noise features are generated from the standard normal distribution. 

\subsection{Comparison with Full Data Estimates}  \label{sec:fullsim}

In this section, we evaluate the ability of the distributed \texttt{MetaLP} approach to consistently estimate the oracle full data LP statistics under various partitioning schemes.  Comparisons are made in terms of variable selection and run time (see Table \ref{tab:tpfp}). Let $n$ be the total number of observations in one dataset, where $n=5,000, 50,000, 500,000, ~\text{and}~ 1 ~ \text{million}$. For each setting of $n$, we generate 100 datasets and randomly partition the dataset with $n$ total observations into $k=\lfloor n^{\gamma}+0.5 \rfloor$ subpopulations with roughly equal numbers of observations, where $\gamma=0.3, 0.4, 0.5$. 

Table \ref{tab:tpfp} provides the mean absolute error ($\times 10^5$) for the \texttt{MetaLP} LP statistic estimates of the oracle full data LP statistics across all 50 variables.  All mean absolute errors are small, indicating estimation using the distributed \texttt{MetaLP} approach is consistent with estimation using the whole dataset. Note that errors increase as the number of partitions, $k$, increase for fixed $n$.  This is expected as the number of observations in each partition decreases as the number of partitions increases for fixed $n$.  However, for fixed $k$, errors are inversely proportional to the number of observations $n$.  Table \ref{tab:tpfp} also compares the \texttt{MetaLP} and oracle full data LP variable selection methods in terms of accuracy in selecting the true model variables, $\{X_1,X_2,X_3\}$, and computation time. Second order LP statistics are used to test for significance for $X_1$, since it has a second order impact on the dependent variable, and first order LP statistics are used to detect significance of other variables.  Note that both methods correctly select all the three important variables every time, which suggests that the distributed approach is comparable to the full data approach in selecting important variables.  However, our distributed \texttt{MetaLP} approach saves a considerable amount of time compared to the non-distributed approach (i.e. computing $\LP$ statistics from the whole dataset all-at-once). We list speed improvements (how many times faster the \texttt{MetaLP} algorithm is over the full data approach) in the last column of Table \ref{tab:tpfp}. For example, when $n=1,000,000$ and $\gamma=0.5$, \texttt{MetaLP} is about \textit{150 fold faster}.


\subsection{Comparison with Other Methods}  \label{sec:BLBsim}

In this section, we compare the performance of our proposed \texttt{MetaLP} framework with two nonparametric, communication-efficient, distributed inference algorithms: BLB \cite{Kleineretal2014} and SAVGM \cite{Zhang2013}. As noted in Table \ref{tab:comparemethods}, BLB and SAVGM provide a way to conduct distributed inference for a given estimator provided by users.  In order to make a fair comparison, we use empirical LP statistic estimators for BLB and SAVGM methods.  We call these methods LP-BLB and LP-SAVGM, respectively, to reflect that they are based on $\LP$ statistics. Similar to Section \ref{sec:fullsim}, we compare the methods based on their abilities to accurately estimate the oracle full data LP statistics, as well as their abilities to differentiate between important and noise variables. We calculate the mean square deviance (MSD) of the distributed LP statistic estimates from the oracle full data LP statistics,
\beq 
\text{MSD} = \frac{1}{R}\sum_{r=1}^{R}  \left \{\widehat{\LP}_r^{*} - \LP^{\text{(full)}}_{r} \right \}^2,  
\eeq
where $R$ is the number of simulated repetitions, $\widehat{\LP}^{*}_r$ are the distributed LP statistic estimates for a specific method, and $\LP^{\text{(full)}}_{r}$ are the oracle full data $\LP$ statistics. 

We use the same model as in Section \ref{sec:fullsim} to generate $R=100$ realizations of the simulated data for each $n=10,000, 50,000, 100,000$ with $\gamma=0.3$ in determining the number of subpopulations for all methods.  For LP-BLB, we set the number of bootstrap samples taken within each subpopulation to be $100$, following \cite{Kleineretal2014}. For LP-SAVGM, we fix the sub-sampling ratio to be $0.08$. The upper portion of Table \ref{tab:blbtable} summarizes the results. 

\begin{table}[ht]
\vskip1em
\begin{center}
\def\arraystretch{1.2}

\resizebox{9cm}{!}{
\begin{tabular}{cc|c|c|c|c|c|c}
\cline{1-7}
\multicolumn{7}{ |c| }{Equal Subpopulation Size}\\ 
\cline{1-7}

\multicolumn{1}{ |c| }{Methods}  & 
\multicolumn{1}{ |c| }{$n$}  & 
\multicolumn{3}{ |c| }{Relative MSD} & 
\multicolumn{1}{ |c| }{Mean} &
\multicolumn{1}{ |c| }{Speed}\\ 

\cline{3-5}
\multicolumn{1}{ |c| } {} & 
\multicolumn{1}{ |c| }{}  & 
\multicolumn{1}{ |c| }{$X_1$} & 
\multicolumn{1}{ |c| }{$X_2$} & 
\multicolumn{1}{ |c| }{$X_3$} & 
\multicolumn{1}{ |c| }{Extra FD} & 
\multicolumn{1}{ |c| }{Increase} \\
\cline{1-7}
\multicolumn{1}{ |c  }{\multirow{1}{*}{} } &
\multicolumn{1}{ |c| }{10,000} & 1.06 & 1.92 & 2.22 & 1.80 & 125\\ 
\multicolumn{1}{ |c  }{\multirow{1}{*}{LP-BLB} } &
\multicolumn{1}{ |c| }{50,000} & 1.06 & 2.68  & 2.98  & 1.75 & 106\\ 
\multicolumn{1}{ |c  }{\multirow{1}{*}{} } &
\multicolumn{1}{ |c| }{100,000} & 1.14  & 3.46 & 3.89 & 1.48  & 97\\ 
\cline{1-7}

\multicolumn{1}{ |c  }{\multirow{1}{*}{} } &
\multicolumn{1}{ |c| }{10,000} &  1.46 & 35.87 & 39.87  & -0.66 & 1.20\\ 
\multicolumn{1}{ |c  }{\multirow{1}{*}{LP-SAVGM} } &
\multicolumn{1}{ |c| }{50,000} & 1.61 & 64.01  & 52.79 &  -0.29 & 1.09\\ 
\multicolumn{1}{ |c  }{\multirow{1}{*}{} } &
\multicolumn{1}{ |c| }{100,000} & 1.67  & 99.97 & 88.61 &  -0.14  & 1.05\\ 
\cline{1-7}
\multicolumn{7}{ |c| }{Unequal Subpopulation Size}\\ 
\cline{1-7}
\multicolumn{1}{ |c| }{Methods}  & 
\multicolumn{1}{ |c| }{$n$}  & 
\multicolumn{3}{ |c| }{Relative MSD} & 
\multicolumn{1}{ |c| }{Mean} &
\multicolumn{1}{ |c| }{Speed}\\ 

\cline{3-5}
\multicolumn{1}{ |c| } {} & 
\multicolumn{1}{ |c| }{}  & 
\multicolumn{1}{ |c| }{$X_1$} & 
\multicolumn{1}{ |c| }{$X_2$} & 
\multicolumn{1}{ |c| }{$X_3$} & 
\multicolumn{1}{ |c| }{Extra FD} & 
\multicolumn{1}{ |c| }{Increase} \\
\cline{1-7}
\multicolumn{1}{ |c  }{\multirow{1}{*}{} } &
\multicolumn{1}{ |c| }{10,000} & 2.42 & 16.61 & 3.45 & 2.22 & 121\\ 
\multicolumn{1}{ |c  }{\multirow{1}{*}{LP-BLB} } &
\multicolumn{1}{ |c| }{50,000} & 8.50 & 241.45  & 4.08  & 2.86 & 102\\ 
\multicolumn{1}{ |c  }{\multirow{1}{*}{} } &
\multicolumn{1}{ |c| }{100,000}  & 12.06 & 585.41 & 6.46 & 4.08  & 96\\ 
\cline{1-7}

\multicolumn{1}{ |c  }{\multirow{1}{*}{} } &
\multicolumn{1}{ |c| }{10,000} &  3.88 & 74.63 & 42.58  & 0.29 & 1.12\\ 
\multicolumn{1}{ |c  }{\multirow{1}{*}{LP-SAVGM} } &
\multicolumn{1}{ |c| }{50,000} & 11.14 & 323.22  & 317.83 &  1.31 & 1.08\\ 
\multicolumn{1}{ |c  }{\multirow{1}{*}{} } &
\multicolumn{1}{ |c| }{100,000} & 23.02  & 727.88 & 909.04 &  1.41  & 1.05\\ 
\cline{1-7}
\end{tabular}
}
\end{center}
\caption{Comparison of methods in estimating full data LP statistics and variable selection.  Relative MSD (e.g. $\mathrm{MSD}_{\text{LP-BLB}}/\mathrm{MSD}_{\text{MetaLP}}$) compares the accuracy in estimating the oracle full data LP statistics.  Mean extra false discovery (FD) is the average number of additional noise variables selected by other methods compared to \texttt{MetaLP}. Speed increase captures how many times faster the \texttt{MetaLP} algorithm runs compare to other methods. Upper portion: under equal subpopulation size; lower portion: under unequal subpopulation size.}  
\label{tab:blbtable}
\end{table}

The relative MSD, $\mathrm{MSD}_{\text{LP-BLB}}/\mathrm{MSD}_{\texttt{MetaLP}}$ and $\mathrm{MSD}_{\text{LP-SAVGM}}/\mathrm{MSD}_{\texttt{MetaLP}}$, are all greater than 1, which means \texttt{MetaLP} is more accurate on average for all three important variables and sample sizes. The LP-BLB method relies on bootstrap resampling to estimate the distribution of the statistic locally.  It has been noted that ``\emph{the bootstrap distribution is an approximate confidence distribution}'' \cite{Efron2013,XieSingh2013}, so there is not much difference in terms of local estimation.  Hence, the improvement in MSD of the \texttt{MetaLP} method over the LP-BLB method largely comes from the different approach to combining inferences.  Rather than weighting each local inference equally, as the LP-BLB method does, \texttt{MetaLP} assigns optimal weights to each local inference adjusting for possible heterogeneity.  As mentioned previously, even under purely random partitioning with equal sample sizes, heterogeneity may exist (see Supplementary Section \ref{sec:titanic}). SAVGM is essentially a bias correction method for divide and recombine estimators. If SAVGM is applied to unbiased estimators, as noted by \cite{Zhang2013}, it could increase the variance of the estimator substantially. This is consistent with our simulation results indicating LP--SAVGM performs significantly worse than \texttt{MetaLP} in terms of MSD. 

In terms of the accuracy in variable selection, all methods correctly select the three important variables on every run. Extra mean false discovery (FD) is the average number of additional noise variables incorrectly determined to be important by the LP-BLB/LP-SAVGM methods compared to the \texttt{MetaLP} method.  For example, when $n=50,000$, the LP-BLB method, on average, falsely selects 1.75 additional noise variables compared to the \texttt{MetaLP} method. It should be noted that LP-SAVGM performs well in terms of selecting fewer noise variables due to the inflated variance of the $\LP$ statistic estimates. 

Computational savings is a crucial consideration for big data analysis.  Note that \texttt{MetaLP} is around 100 times faster than LP-BLB.  The additional LP-BLB run time comes from the need to resample each subpopulation numerous times in order to obtain the bootstrap estimate, while our approach calculates the LP estimates for each subpopulation in one shot.  The LP-SAVGM is relatively comparable to \texttt{MetaLP}, where the additional run time is due to the need for subsampling to perform the bias correction.

Next, we investigate the impacts of heterogeneity on the different methods. For this exercise, we will define heterogeneity in terms of varying subpopulation sizes, letting the size increase linearly. In particular, we set the size of the first subpopulation to be 500 and increase the size by 150 for the second subpopulation, and so on. The lower portion of Table \ref{tab:blbtable} shows that the relative MSD increases dramatically, indicating that heterogeneity has substantial negative impacts on both LP-BLB and LP-SAVGM, while \texttt{MetaLP} remains robust under this setting.  It also should be noted that, unlike in the equal size case, \texttt{MetaLP} tends to outperform LP-SAVGM in terms of mean extra false discoveries, especially when the total sample size is large.

\section{Expedia Personalized Hotel Search Dataset}\label{sec:expedia}

Based on \texttt{MetaLP}, in this section we develop a model-free, parallelizable, two-sample feature selection algorithm for big data and apply it to the Expedia digital marketing problem. Detailed discussions on each of the following components of our big data two-sample inference model are given in the next sections:
\begin{itemize}
  \item (Section \ref{sec:data}) Data Description.
  \item (Section \ref{sec:partition}) Data Partitioning.
  \item (Section \ref{sec:LP-map}) LP Map Function.
  \item (Section \ref{sec:heter}) Heterogeneity Diagnostic and Regularization.
  \item (Section \ref{sec:reduce}) Meta Reducer via LP Confidence Distribution.
  \item (Section \ref{sec:robust}) Robustness to Size and Number of Subpopulations.
\end{itemize}
\subsection{Data Description}
\label{sec:data}

\begin{figure}[ht] 
\centering
\includegraphics[scale=.27,trim=.1cm 0cm .1cm 0cm]{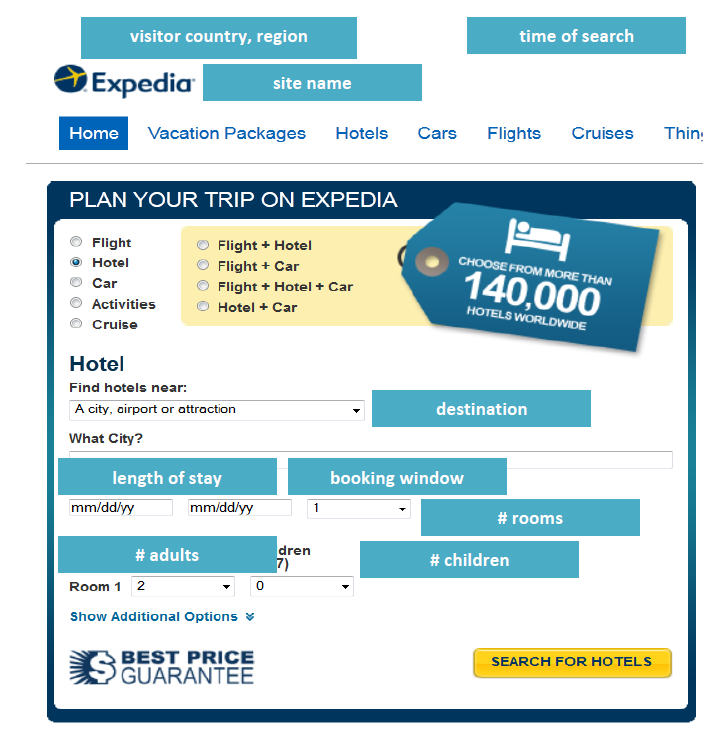}\\
\vspace{0.2cm}
\includegraphics[scale=.27,trim=.1cm 0cm .1cm 0cm]{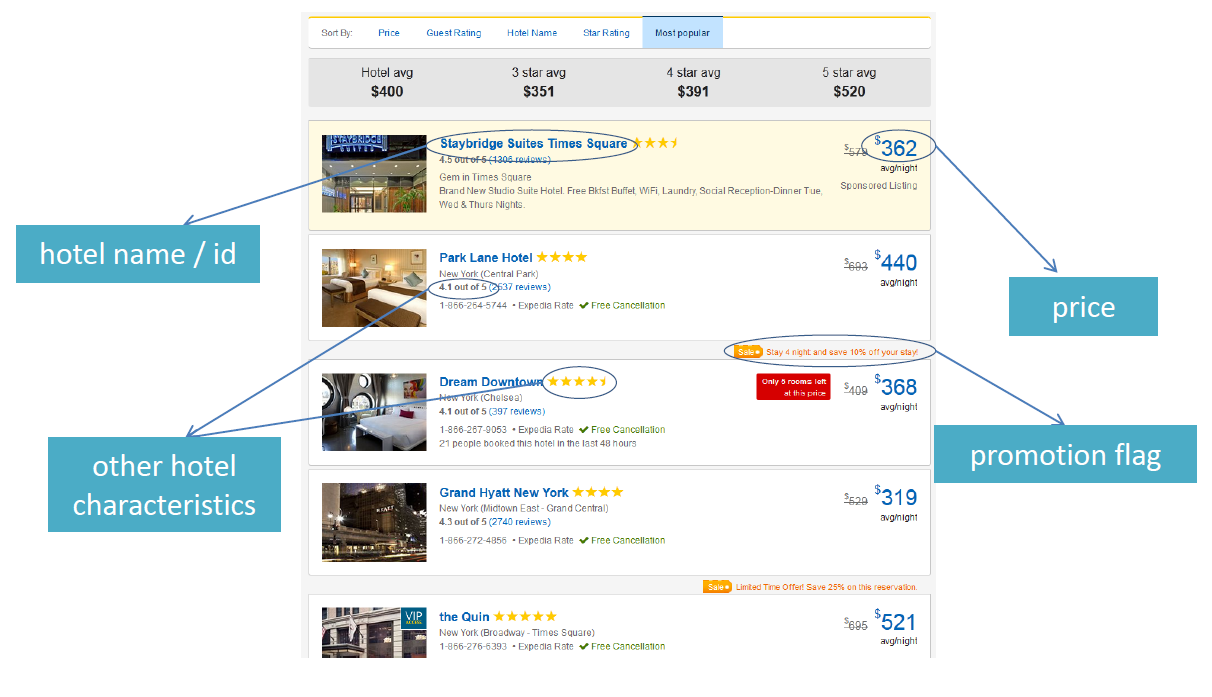} 
\caption{On top (a) is a snapshot of a search window with search criteria variables; on the bottom (b) is a list of ranked hotels returned by Expedia with hotel characteristic variables.}
 \label{fig:hotel}
\end{figure} 

Expedia provided a large dataset ($n=9,917,530$) of hotel search results collected over a window of the year $2013$ \cite{KaggleData} in order to better understand the factors that influence booking behavior.  Data are generated by online customers who first provide search criteria for their desired travel plans (e.g. length of stay, destination, number of children, etc.) to the Expedia website (see Figure \ref{fig:hotel}a).  Expedia then returns an ordered list of available hotels along with important hotel information (e.g. hotel name, price, star rating, promotion, etc.) for customers to review and consider booking for their travel plans (see Figure \ref{fig:hotel}b).  In the background, Expedia also records important user information (e.g. visitor location, search history, etc.) and competitor pricing and availability for hotels listed, which may impact booking behavior.  Expedia then records how customers interact with each hotel listed (e.g. ignored, clicked, booked). We are primarily interested in understanding which factors influence the binary response variable, \texttt{booking\_bool}, indicating whether the hotel was booked or not.  Descriptions of representative variables and data types are provided in Supplementary Section \ref{sec:ExpediaDataDes}.

\subsection{Data Partitioning}
\label{sec:partition}
We consider two different partitioning schemes that are appropriate for the Expedia dataset: 1) random partitioning, which results in homogeneous, similarly sized subpopulations, and 2) predefined partitioning, which results in heterogeneous, disproportionately sized subpopulations.

{\bf Step 1.} We randomly assign search lists, which are collections of observations with the same search id in the dataset, to $200$ different subpopulations.  Random assignment of search lists rather than individual observations ensures that sets of hotels viewed in the same search session are all contained in the same subpopulation. Note that the number of subpopulations chosen can be adapted to meet the processing and time requirements of different users.  We show in Section \ref{sec:robust} that our method is robust to different numbers of subpopulations as the inference remains unchanged. 

There may be situations where natural groupings exist in the dataset, which can be directly used to form subpopulations. For example, the available Expedia data could be grouped naturally by the country where each visitor to the Expedia website resides, \texttt{visitor\_location\_country\_id}.

\begin{figure}[ht]
 \centering
 \includegraphics [scale=.4]{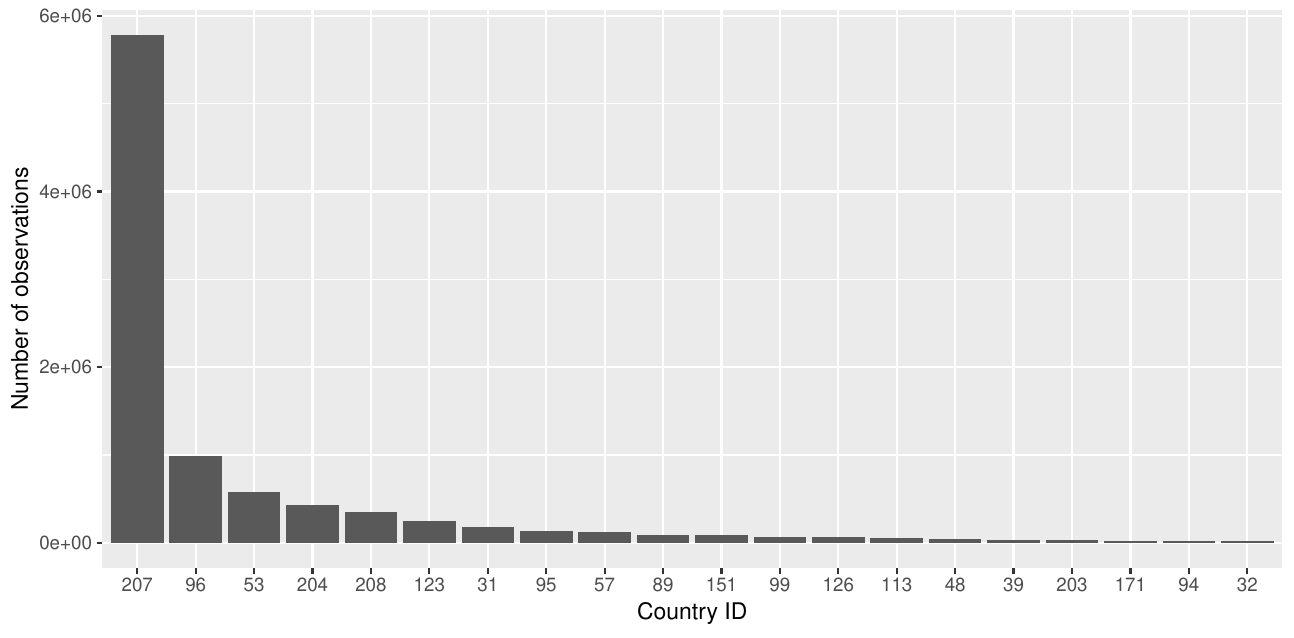}
  \caption{ \small Number of observations for the 20 largest subpopulations from partitioning by \texttt{visitor\_location\_country\_id}}
  \label {fig:bar}
\end{figure}

Our framework can directly utilize these predetermined subpopulations for processing rather than requiring the massive data to be gathered and randomly assigned to subpopulations.  However, this partitioning scheme may result in heterogeneous subpopulations, so extra steps must be taken to address this issue as described in Section \ref{sec:heter}. For the Expedia dataset, Figure \ref{fig:bar} shows the number of observations for the 20 largest subpopulations from partitioning by \texttt{visitor\_location\_country\_id}. The top three largest countries by number of observations contain 74\% of the total observations, and the leading country contains almost 50\% of the total observations. On the other hand, random partitioning results in roughly equal sample sizes across subpopulations (50,000 each).

\subsection{LP Map Function}
\label{sec:LP-map}
We tackle the data variety problem by developing automated mixed data algorithms using LP statistical data modeling tools. 

{\bf Step 2.} Following the theory in Section \ref{sec:LP}, we construct LP score polynomials, $T_{j}(x;X_i)$, for each variable based on each partitioned input dataset. Figure \ref{fig:score} shows the LP basis polynomials for variables \texttt{variable\_length\_of\_stay} (discrete) and \texttt{price\_usd} (continuous).

{\bf Step 3.} Estimate $\LP_{\ell}[j;X_i,Y]$, which denotes the $j$th LP statistic for the $i$th variable in the $\ell$th subpopulation, 
\beq \widetilde{\LP}_{\ell}[j;X_i,Y]\,=\, n_{\ell}^{-1}\sum_{k=1}^{n_{\ell}} T_j(x_k;X_i)\,T_1(y_k;Y).\eeq

{\bf Step 4.} Compute the corresponding LP confidence distribution given by 
\beq \Phi\left( \sqrt{n}\left(\LP_{\ell}[j;X_i,Y]-\widehat{\LP}_{\ell}[j;X_i,Y]\right)\right), \eeq

for $i=1,\ldots,45$ variables across $\ell=1, \ldots, 200$ random subpopulations (or $233$ predefined subpopulations defined by \texttt{visitor\_location\_country\_id}).

\subsection{Heterogeneity Diagnostic and Regularization}
\label{sec:heter}
\begin{figure}[ht]
 \centering
\hspace{-0.25cm}\includegraphics [scale=.45]{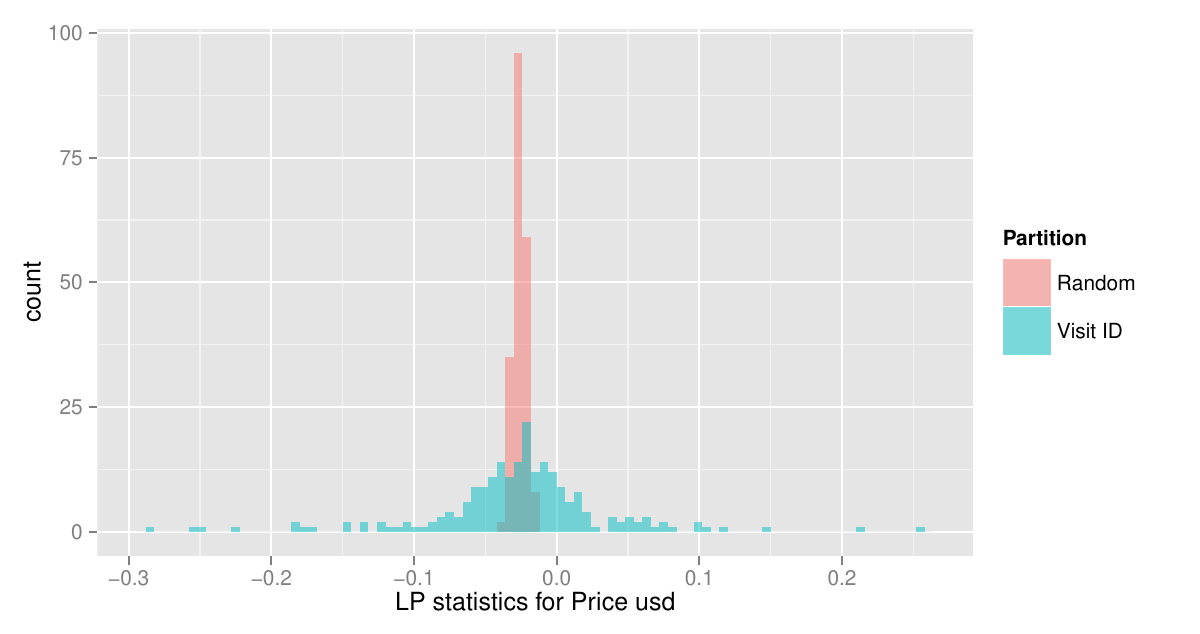}
  \caption{Distribution of LP statistic estimates for the variable \texttt{price\_usd} based on random partitioning and \texttt{visitor\_location\_country\_id} partitioning.}
  \label {fig:LP}
\end{figure}

Figure \ref{fig:LP} shows the distribution of the first order LP statistic estimates for variable \texttt{price\_usd} across different subpopulations based on random and \texttt{visitor\_location\_country\_id} partitioning. It is clear that random partitioning produces relatively homogeneous LP statistic estimates as the distribution is much more concentrated.  On the other hand, \texttt{visitor\_location\_country\_id} partitioning results in heterogeneous LP statistic estimates, which is reflected in the dispersion of the corresponding histogram. In fact, the standard deviation of the first order LP statistic under \texttt{visitor\_location\_country\_id} partitioning is about $15$ times more than that of the random partition, which further highlights the underlying heterogeneity issue. Thus, care must be taken to account for this heterogeneity in a judicious manner that ensures consistent inference. We advocate the method mentioned in Section \ref{sec:I2}.

\begin{figure}[th] 
\centering
\vspace{-0.25cm}
\hspace{-1.75cm}\includegraphics[height=5cm,width=6.6cm]{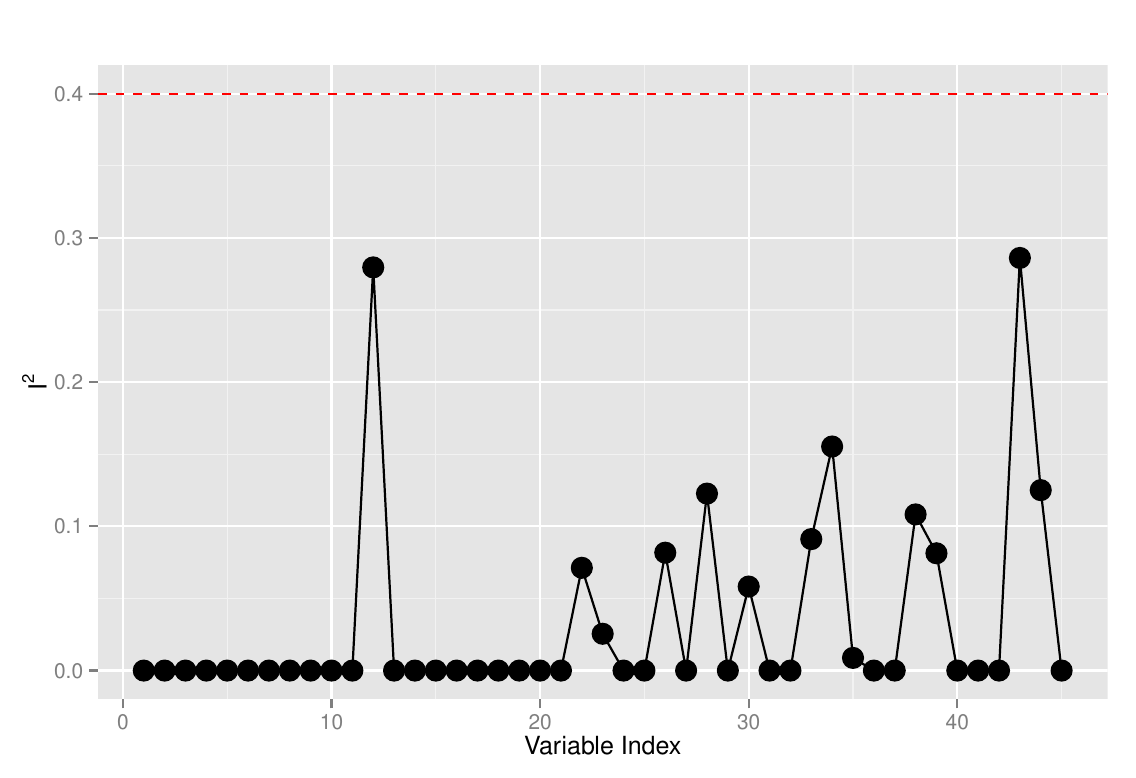}\\
\vspace{-1cm}
\hspace{-0.25cm}\includegraphics[height=5cm,width=9.25cm]{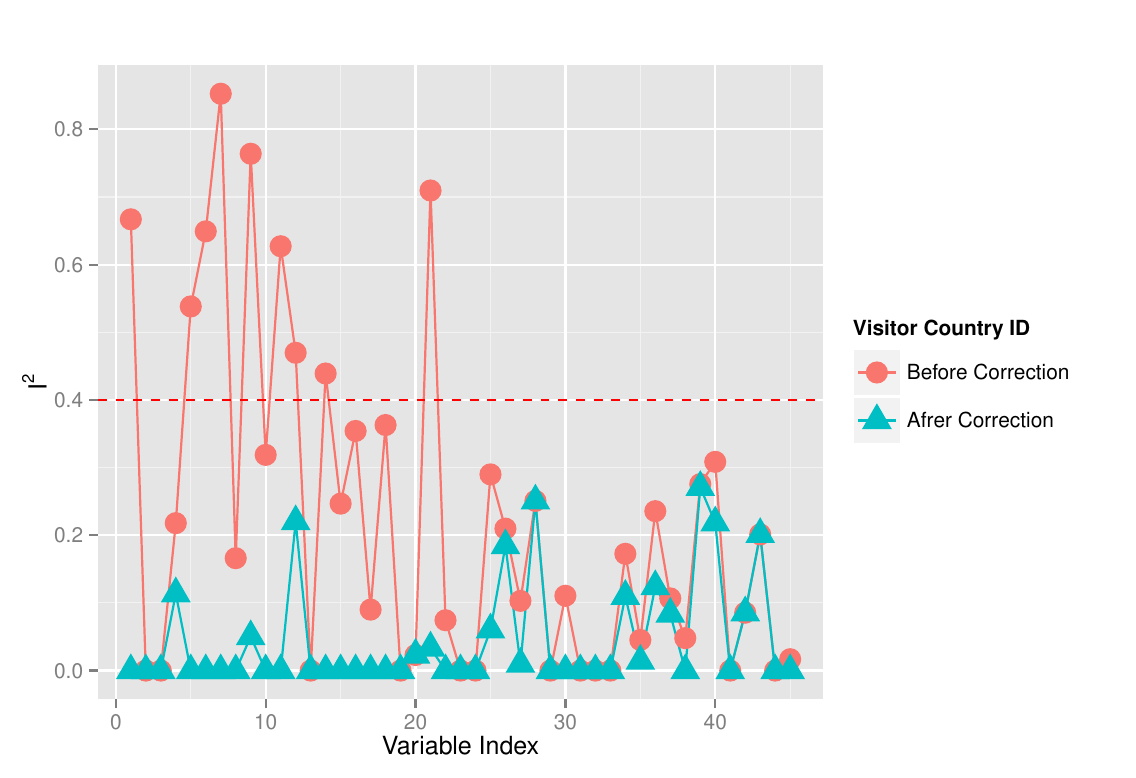}
 \caption{On top (a) is a plot of the $I^2$ Diagnostic under random partitioning; on bottom (b) is a comparison of the $I^2$ diagnostic for the \texttt{visitor\_location\_country\_id} partitioning before $\tau^2$ correction and after $\tau^2$ correction.}
 \label {fig:sitei}
\end{figure}

{\bf Step 5.} Compute the Cochran's Q-statistic using \eqref{eq:Q} and $I^2$ heterogeneity index \eqref{eq:I2} based on $\LP_1[j;X_i,Y],\ldots, \LP_k[j;X_i,Y]$ for each $i$ and $j$, where $k$ is the number of subpopulations. Under random partitioning, the subpopulations are fairly homogeneous, with respect to all variables, as all $I^2$ statistics are below 40\% (see Figure \ref{fig:sitei}(a)). However, \texttt{visitor\_location\_country\_id} partitioning divides data into heterogeneous subpopulations for some variables as shown in Figure \ref{fig:sitei}(b) (i.e. some variables have $I^2$ values outside the permissible range of 0 to 40\% before correction). 

{\bf Step 6.} Compute the DerSimonian and Laird data-driven estimate
\[\hat{\tau}_i^2 = \max \left\{
0,
\frac{Q_i-(k-1)}
{n - \sum_\ell  n_\ell ^{2} / n }
\right\},~~~~~i=1,\ldots,p.~~~\]
One can also use other enhanced estimators, like the restricted maximum-likelihood estimator, as discussed in Supplementary Section \ref{sec:tau2estimator}. $I^2$ diagnostics \textit{after} correction using $\tau^2$ regularization are shown in Figure \ref{fig:sitei}(b).  Note that all $I^2$ values after correction fall within the acceptable range of 0 to 40\%. This result demonstrates that our framework can resolve heterogeneity issues among subpopulations through $\tau^2$ regularization, which protects the validity of the meta-analysis approach.

\subsection{Meta Reducer via LP Confidence Distribution}
\label{sec:reduce}
This step combines confidence distribution estimates of $\LP$ statistics from different subpopulations to estimate the combined confidence distribution of the $\LP$ statistic for each variable as outlined in Section \ref{sec:htau}. 

{\bf Step 7.} Use $\tau^2$-corrected weights to  properly take into account the  heterogeneity effect. Compute $\widehat{\LP}^{(c)}[j;X,Y])$ by \eqref{eq:taulp} and the corresponding LP confidence distribution using Theorem \ref{thm:taulp}.

\begin{figure}[!tth] 
\centering
\includegraphics[scale=.44,trim=.5cm .5cm .5cm .8cm]{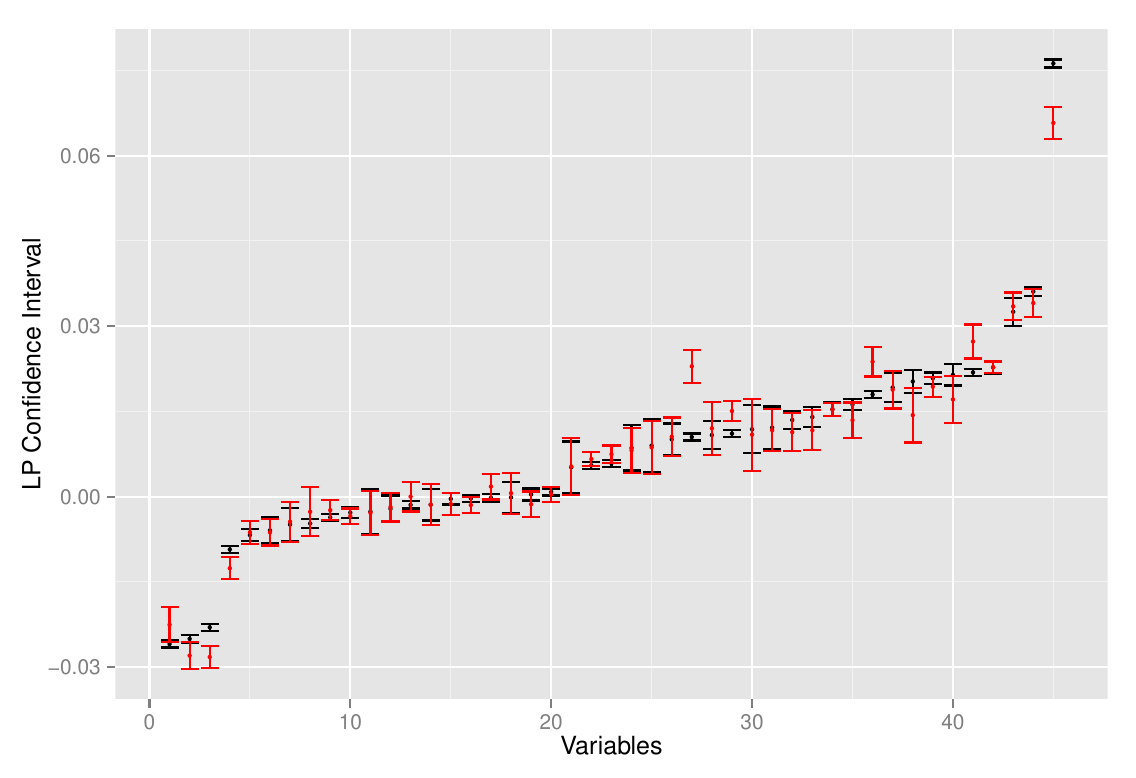}
 \caption{95\% Confidence intervals for $\LP$ statistics for each variable in the Expedia dataset under random partitioning (black) and \texttt{visitor\_location\_country\_id} partitioning (red).}
 \label{fig:sitelp}
\end{figure}

The resulting 95\% confidence intervals for each variable under both random and \texttt{visitor\_location\_country\_id} partitioning can be found in Figure \ref{fig:sitelp}.  Variables with indexes $43$, $44$, and $45$ have highly significant positive relationships with \texttt{booking\_bool}, the binary response variable. Those variables are \texttt{prop\_location\_score2}, the second score quantifying the desirability of a hotel's location, \texttt{promotion\_flag}, if the hotel had a sale price promotion specifically displayed, and \texttt{srch\_query\_affinity\_score}, the log probability a hotel will be clicked on in Internet searches. There are three variables that have highly negative impacts on hotel booking: \texttt{price\_usd}, displayed price of the hotel for the given search, \texttt{srch\_length\_of\_stay}, number of nights stay that was searched, and \texttt{srch\_booking\_window}, number of days in the future the hotel stay started from the search date. Moreover, there are several variables whose $\LP$ statistic confidence intervals include zero, which means those variables have an insignificant influence on hotel booking. The top five most influential variables in terms of absolute value of LP statistic point estimates are \texttt{prop\_location\_score2}, \texttt{promotion flag},   \texttt{srch\_query\_affinity\_score}, \texttt{price\_usd}, and \texttt{srch\_length\_of\_stay} (see Table \ref{tab:top}). Intuitively, users are more likely to book hotels with desirable locations (high \texttt{prop\_location\_score2} values), special promotions (\texttt{promotion\_flag}=1), and high probabilities of being clicked (high \texttt{srch\_query\_affinity\_score} values). The variables we selected are also among the list of top important variables identified by the winners of the ICDM 2013 competition \cite{kaggleWinner}, which required participants to develop hotel ranking algorithms for all user search queries based on the features in the Expedia dataset.  This speaks to the usefulness of these selected features for downstream analytical tasks (e.g. classification, ranking, etc.). 

Note that the confidence intervals for each of the variables under both partitioning schemes are very similar, resulting in similar variable selection outcomes.  Four of the top five influential variables identified under random partitioning are also in the top five influential variables identified under \texttt{visitor\_location\_country\_id} partitioning (see Table \ref{tab:top}).  The impact of heterogeneity on the results under \texttt{visitor\_location\_country\_id} partitioning can be seen in Figure \ref{fig:sitelp} as the confidence intervals are generally wider than those derived under random partitioning.  This can be attributed to extra variability among subpopulations captured by $\tau^2$ due to different characteristics among subpopulations defined by country.

\begin{table}[ht]
\begin{center}
\resizebox{8.8cm}{!}{
    \begin{tabular}{| c | l | l |}
    \hline
    Rank & Random partition & Predetermined partition  \\ \hline
    1 & \texttt{prop\_location\_score2} & \texttt{prop\_location\_score2}  \\ \hline
    2 & \texttt{promotion\_flag} & \texttt{promotion\_flag}  \\ \hline
    3 & \texttt{srch\_query\_affinity\_score} & \texttt{srch\_query\_affinity\_score} \\
    \hline
    4 & \texttt{price\_usd} & \texttt{srch\_length\_of\_stay}\\
    \hline
    5 & \texttt{srch\_length\_of\_stay} & \texttt{srch\_booking\_window} \\
    \hline
    \end{tabular}
    }
\end{center}
\caption{Top five influential variables by random partitioning and predetermined partition}
\label{tab:top}
\end{table}

\subsection{Robustness to Size and Number of Subpopulations} \label{sec:robust}
\begin{figure}[ht]
\centering
\includegraphics [scale=.35]{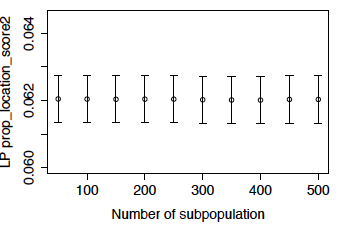}
\includegraphics [scale=.35]{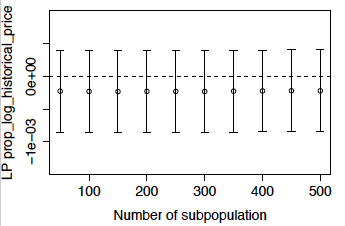}
\includegraphics [scale=.35]{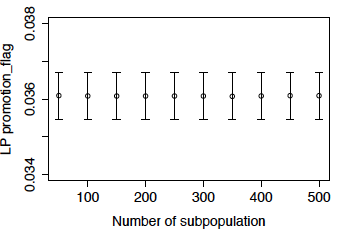}
\includegraphics [scale=.35]{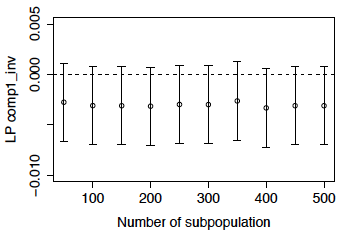}
\includegraphics [scale=.35]{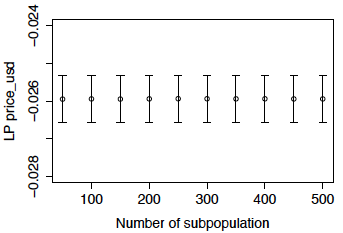}
\includegraphics [scale=.35]{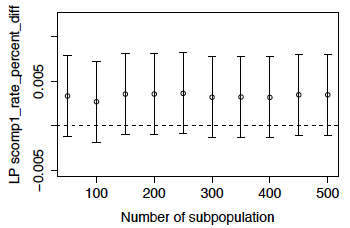}
\caption{LP statistics and 95\% confidence intervals for six variables across different numbers of subpopulations (dotted line is at zero).}
  \label {fig:sub}
\end{figure}

Due to different capabilities of computing systems available to users, users may choose different sizes and numbers of subpopulations for distributed computing. This requires our algorithm to be robust to different numbers and sizes of subpopulations for practical applications.  To assess robustness, we compare LP statistic estimates generated from multiple random partitions with different numbers of subpopulations ($k=50,100,150,200,250,300,350,400,450,500$) for the Expedia dataset.  Figure \ref{fig:sub} presents LP statistic 95\% confidence intervals for three influential variables and three insignificant variables calculated from partitions with varying numbers of subpopulations.  Note that the intervals are consistent, even as the number of subpopulations increase (i.e. the number of observations in each subpopulation decrease), which is evidence of stable estimation.

\section{Final Remarks}
\label{sec:conclusion}
To address the major challenges associated with big data analysis, we have outlined a general theoretical foundation in this article, which we believe may provide the missing link between small data and big data science. Our research shows how the traditional and modern `small' data modeling tools can be successfully adapted and connected for developing powerful, big data analytic tools by leveraging distributed computing environments. 

In particular, we have proposed a nonparametric two sample inference algorithm that has the following two-fold practical significance for solving real-world data mining problems: (1) scalability for large data by exploiting distributed computing architectures using a confidence distribution based meta-analysis framework, and (2) automation for mixed data using a united LP computing formula. Undoubtedly, our theory can be adapted for other common data mining problems, and we are currently investigating how the proposed framework can be utilized to develop parallelizable regression and classification algorithms for big data. 

Instead of developing distributed versions of statistical algorithms on a case-by-case basis, here we develop a generic platform to extend traditional and modern statistical modeling tools to large datasets using scalable, distributed algorithms.  We believe this research is a great stepping stone towards developing a United Statistical Algorithm \cite{ParzenMukhopadhyay2013USM} to bridge the increasing gap between the theory and practice of small and big data analysis.

\ifCLASSOPTIONcompsoc
  \section*{Acknowledgments}
\else

\section*{Acknowledgment}
\fi

SM thanks William S. Cleveland for pointing out relevant literature and providing helpful comments. We also thank the editor, associate editor, and two anonymous reviewers for their constructive comments, which helped us to improve the clarity and organization of the manuscript. This research is supported by the Fox Young Scholars grant, Fox School PhD Student Research Award, and Best Paper Award in the ASA Section on Nonparametric Statistics, JSM 2016. 

\ifCLASSOPTIONcaptionsoff
  \newpage
\fi



%

\bibliographystyle{IEEEtran}
\bibliography{IEEEabrv,mybib}
\vspace{-.8cm}
%

\vspace{-0.3cm}
\begin{IEEEbiography}[{\includegraphics[width=1in,height=1.25in,clip,keepaspectratio]{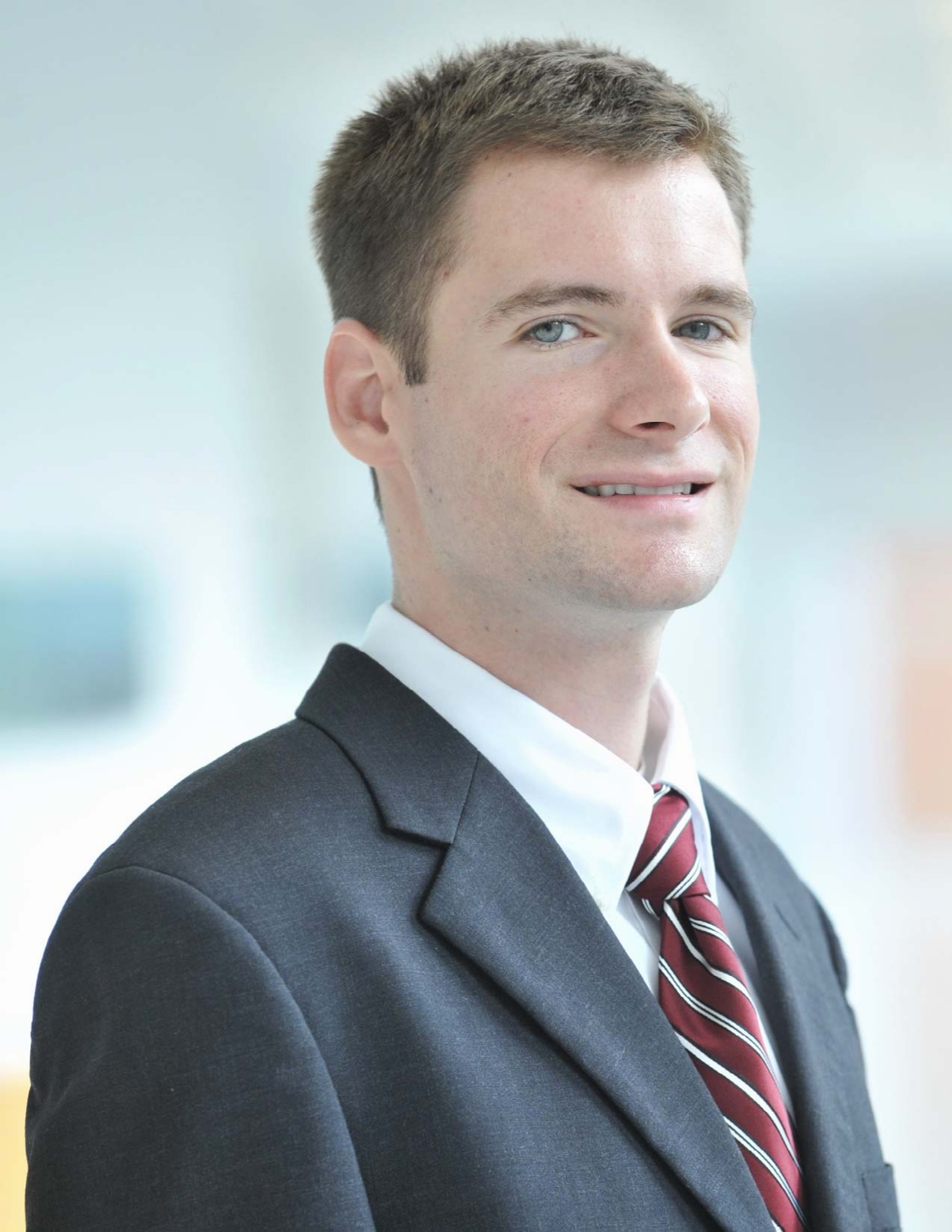}}]%
{Scott Bruce} received his B.S. in Industrial and Systems Engineering and M.S. in Statistics from the Georgia Institute of Technology, USA in 2009 and 2010.  He is currently a Ph.D. candidate in Statistics at Temple University, USA.  His research interests include nonstationary time series analysis, big data learning, Bayesian data analysis, and nonparametric statistics with applications in medicine, public health, finance, and sports.
\end{IEEEbiography}

\begin{IEEEbiography}[{\includegraphics[width=1in,height=1.25in,clip,keepaspectratio]{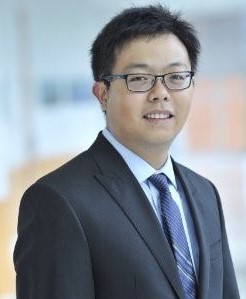}}]%
{Zeda Li}
received his B.S. in Electrical Engineering from Central South University of Forestry and Technology, China in 2010 and M.S. in Biostatistics from Middle Tennessee State University, USA in 2013. He is currently a Ph.D. candidates in Statistics at Temple University, USA. His main research interests are time series analysis, big data analytics, high--dimensional statistics, sufficient dimension reduction methods, and functional data analysis.
\end{IEEEbiography}

\begin{IEEEbiography}
[{\includegraphics[width=1in,height=1.25in,clip,keepaspectratio]{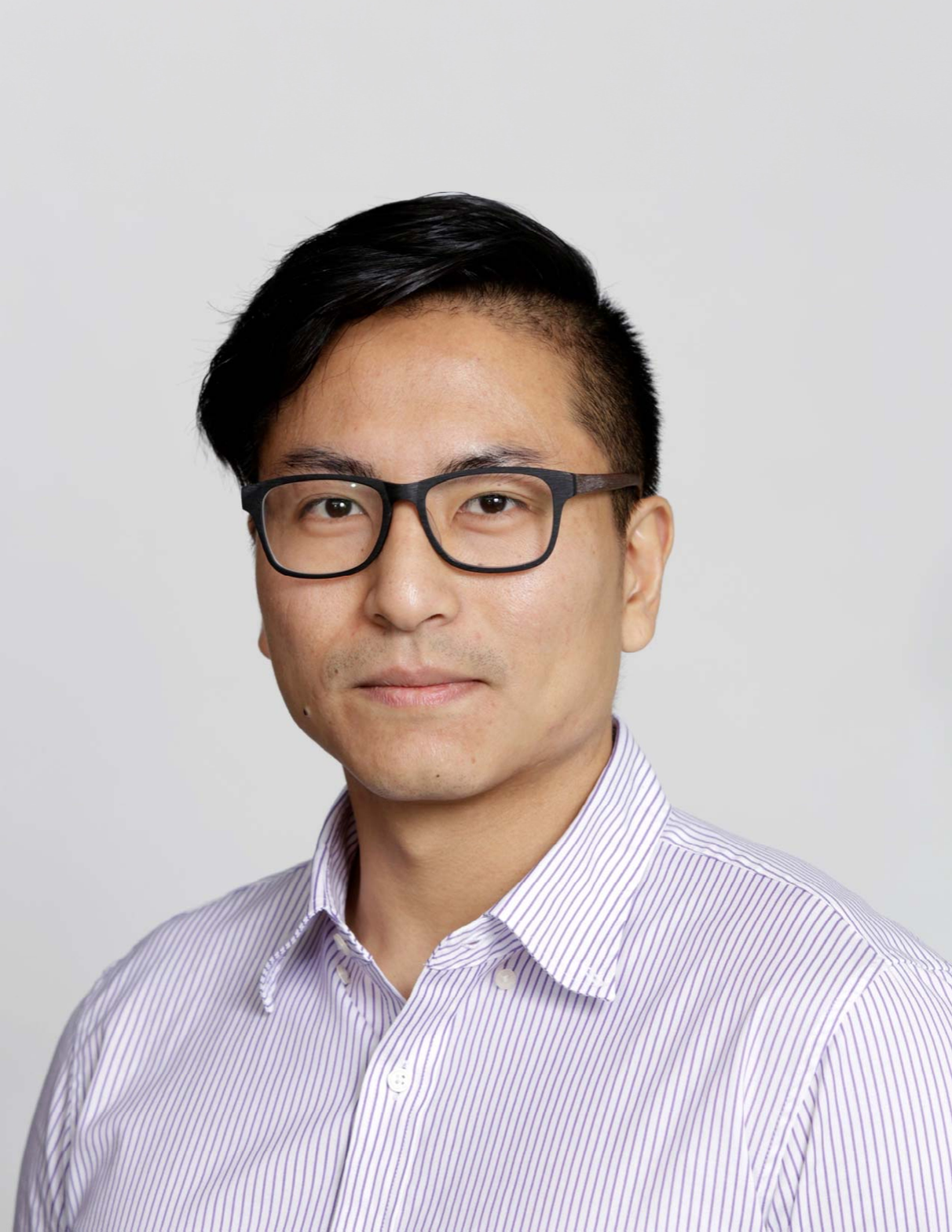}}]%
{Hsiang-Chieh Yang} received his B.A. in Public Finance at National Chengchi University, Taiwan and M.S. in Statistics at Temple University, USA. He is currently a Ph.D. student in Accounting at the University of British Columbia, Canada.  His major research interests include the relationship between financial reporting transparency and accounting standards, internal controls in financial institutions, and applications of machine learning in accounting research.
\end{IEEEbiography}

\begin{IEEEbiography}[{\includegraphics[width=1in,height=1.25in,clip,keepaspectratio]{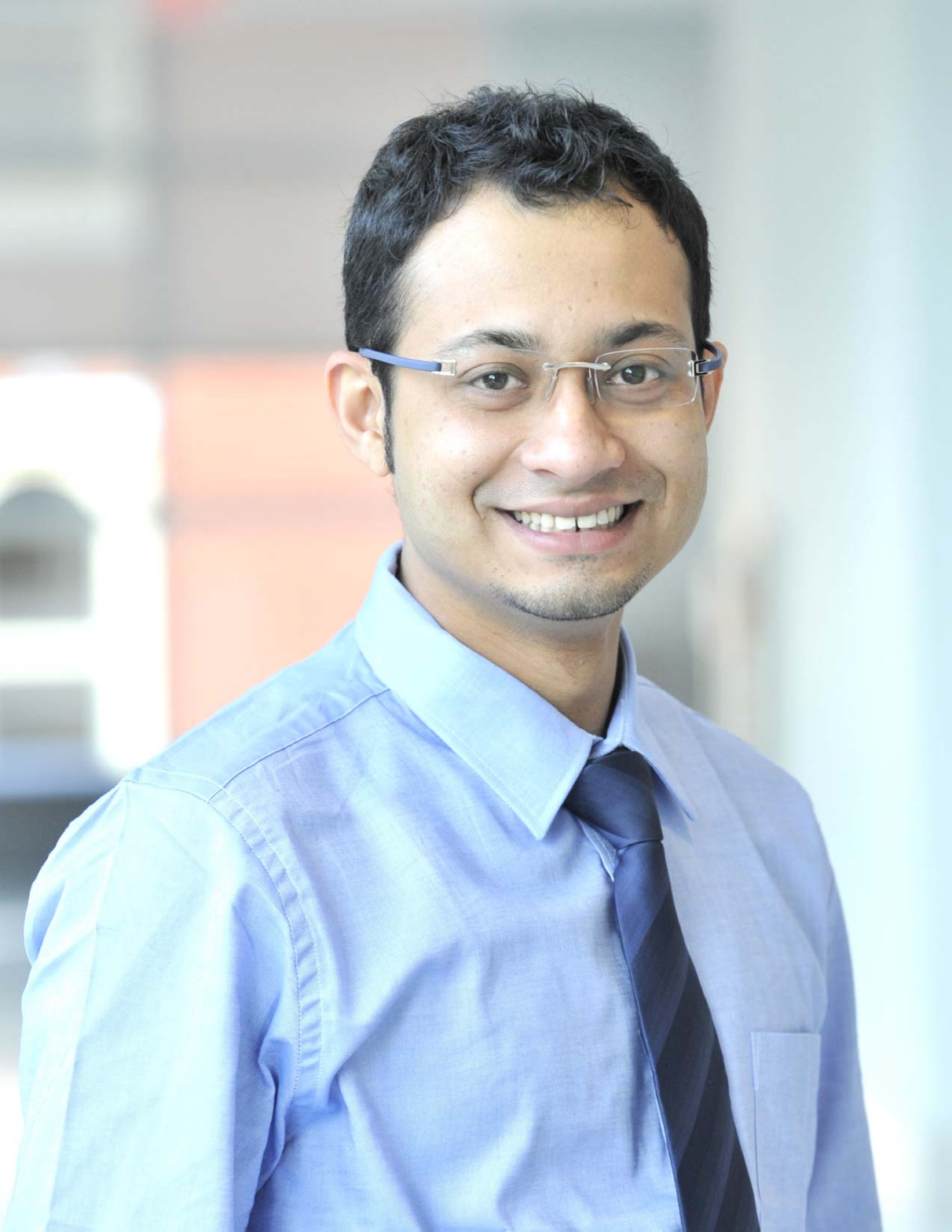}}]%
{Subhadeep Mukhopadhyay}
received his Ph.D. degree in Statistics from Texas A\&M University, USA in 2013. He is currently an Assistant professor of Statistical Science at Fox Business School, Temple University. He is also a member of the Data Science Institute of Temple University. His major research interest lies in developing theory of ``United Statistical Algorithms'' that could reveal the interconnectedness among different branches of statistics. He is a member of the IEEE.
\end{IEEEbiography}

\vfill








\clearpage

\appendices

\vspace{0.5cm}
\setcounter{figure}{0}
\setcounter{table}{0}
\setcounter{equation}{0}
\begin{center} {\Large \bf{The Supplementary Appendix}} \end{center} 

This supplementary document contains five Appendices. Section \ref{sec:ExpediaDataDes} provides a data dictionary for representative variables from various categories found in the Expedia dataset.  Section \ref{sec:titanic} provides a small data example to demonstrate the applicability of the \texttt{MetaLP} framework on datasets both big and small.  Section \ref{sec:paradox} demonstrates how the proper treatment of heterogeneity through the \texttt{MetaLP} approach provides new insights and resolutions for two challenging problems: Simpson's paradox and Stein's paradox.  Finally, Sections \ref{sec:mapreducecode} and \ref{sec:tau2estimator} will describe a \texttt{MapReduce} computational implementation of the MetaLP inference engine and the $\tau^2$ estimators used in our calculations. 

\section{Expedia Data Dictionary}\label{sec:ExpediaDataDes}
See Table \ref{tab:des} for detailed descriptions of representative variables from each category of data found in the Expedia dataset.  Data type information is also included to better illustrate the challenges stemming from the mixed data problem.
\begin{table*}[hbp]
\vspace*{\fill}
\centering
\def\arraystretch{1.2}
\begin{tabular}{ |l|l|l|l| }
\hline
Category & Variable & Data Type & Description \\ \hline
\multirow{4}{1cm}{User\\[.2\baselineskip]
        information}  & \texttt{visitor\_location\_country\_id }  & Discrete & The ID of the country in which customer is located \\
 & \texttt{visitor\_hist\_starrating} & Continuous & The mean star rating of hotels customer previously purchased \\
 & \texttt{visitor\_hist\_adr\_usd } & Continuous & The mean price of hotels customer previously purchased\\
 & \texttt{orig\_destination\_distance } & Continuous & Physical distance between hotel and the customer\\ 
 \hline
\multirow{6}{1cm}{Search\\[.2\baselineskip]
        criteria} 
 & \texttt{srch\_length\_of\_stay }  & Discrete & Number of nights stay searched \\
 & \texttt{srch\_booking\_window} & Discrete & Number of days in the future the hotel stay started \\
 & \texttt{srch\_adults\_count} & Discrete & Number of adults specified in the hotel room\\
 & \texttt{srch\_children\_count} & Discrete & Number of children specified in the hotel room\\ 
 & \texttt{srch\_room\_count} & Discrete & Number of hotel rooms specified in the search\\ 
 & \texttt{srch\_saturday\_night\_bool} & Binary & Short stay including Saturday night \\  
\hline
\multirow{7}{1cm}{Static\\[.2\baselineskip]
        hotel\\[.2\baselineskip]
        characteristics} 
 & \texttt{prop\_country\_id }  & Discrete & Country ID where customer is located \\
 & \texttt{prop\_starrating} & Discrete & Hotel star rating \\
 & \texttt{prop\_review\_score} & Continuous & Mean hotel customer review score \\
 & \texttt{prop\_location\_score1} & Continuous & Desirability of hotel location (1)\\ 
 & \texttt{prop\_location\_score2} & Continuous & Desirability of hotel location (2)\\ 
 & \texttt{prop\_log\_historical\_price} & Continuous & Mean hotel price over last trading period \\  
 & \texttt{pprop\_brand\_bool} & Discrete & Independent or belongs to a hotel chain \\  
\hline
\multirow{3}{2cm}{Dynamic\\[.2\baselineskip]
        hotel\\[.2\baselineskip]
        characteristics} 
 & \texttt{price\_usd }  & Continuous & Displayed hotel price for the given search \\
 & \texttt{promotion\_flag} & Discrete & Hotel sale price promotion available  \\
 & \texttt{gross\_booking\_usd} & Continuous & Total transaction value \\
\hline 
\multirow{3}{1cm}{Competitor\\[.2\baselineskip]
        information} 
 & \texttt{comp1\_rate\_percent\_dif }  & Continuous & Absolute percentage difference between competitors  \\
 & \texttt{comp1\_inv } & Binary & If competitor 1 has hotel availability  \\
 & \texttt{comp1\_rate} & Discrete & If Expedia has lower/same/higher price than competitor \\
\hline 
\multirow{2}{1cm}{Other\\[.2\baselineskip]
        information} 
 & \texttt{srch\_id }  & Discrete & Search ID \\
 & \texttt{site\_id }  & Discrete & Expedia Point of Sale ID  \\
\hline 
\multirow{1}{*}{Response } 
 & \texttt{booking\_bool}  & Binary & Hotel booked or not  \\
\hline 
\end{tabular}
\caption{Data dictionary for Expedia dataset.}
\label{tab:des}
\end{table*}


\section{MetaLP Analysis of Titanic Data}\label{sec:titanic}

The {\it Titanic} dataset is utilized as a benchmark to validate the effectiveness, accuracy, and robustness of the \texttt{MetaLP} analytical framework. Due to its manageable size, we are able to compute the full data LP estimates and can compare with the \texttt{MetaLP} estimates, which operate under a distributed computing framework. A step-by-step \texttt{MetaLP} analysis of Titanic dataset is provided here.

The {\it Titanic} dataset contains information on $891$ of its passengers, including which passengers survived.  A key objective in analyzing this dataset is to better understand which factors (e.g. age, gender, class, etc.) significantly influence passenger survival.  Complete descriptions of all eight variables can be found in Table \ref{tab:Tit.var}.  We seek to estimate the relationship between various passenger characteristics ($X_i, i=1,\ldots,7$) and the binary response variable ($Y$), passenger survival, by using both our distributed algorithm and traditional aggregated $\LP$ statistics to compare their results. 

 \begin{figure}[!htb]
  \centering
 \includegraphics[scale=0.4,keepaspectratio,trim=2.5cm .25cm 2.5cm .5cm]{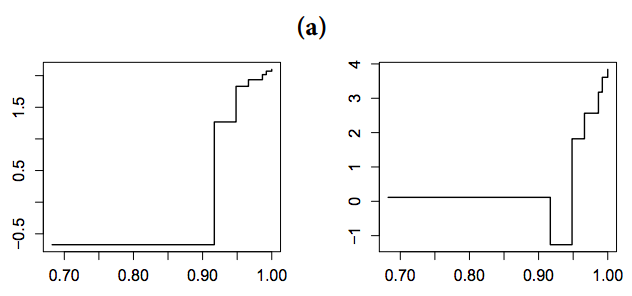}\\
 \vspace{.5cm}
  \includegraphics[scale=0.4,keepaspectratio,trim=2.5cm .25cm 2.5cm .5cm]{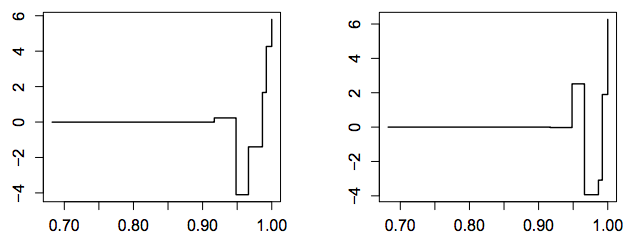}\\
  \vspace{.5cm}
  \includegraphics[scale=0.4,keepaspectratio,trim=2.5cm .25cm 2.5cm .5cm]{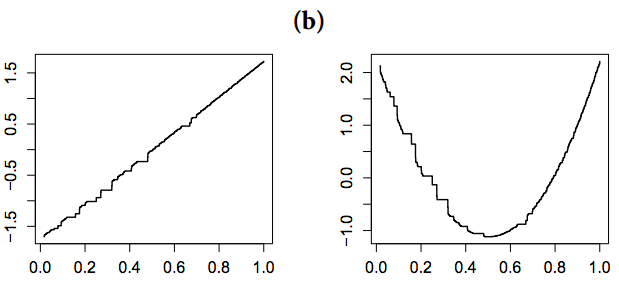}\\
  \vspace{.5cm}
  \includegraphics[scale=0.4,keepaspectratio,trim=2.5cm .25cm 2.5cm .5cm]{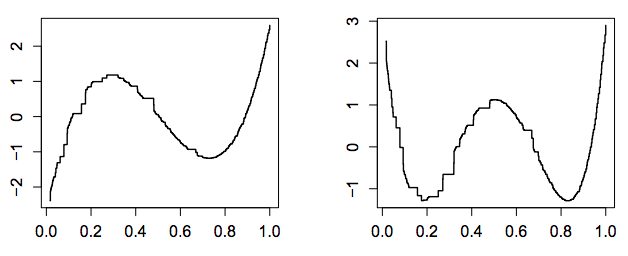}

 \caption{(a) Top: first four LP orthonormal score functions for variable \texttt{\# Siblings/Spouses Aboard}, a discrete random variable taking values $0,\ldots,8$; (b) Bottom: first four LP orthonormal score functions for continuous variable \texttt{Passenger Fare}.} \label{fig:SF}
 \end{figure}
 
 \begin{figure} [ht!] 
\centering
\includegraphics[height=2.3in,width=4in]{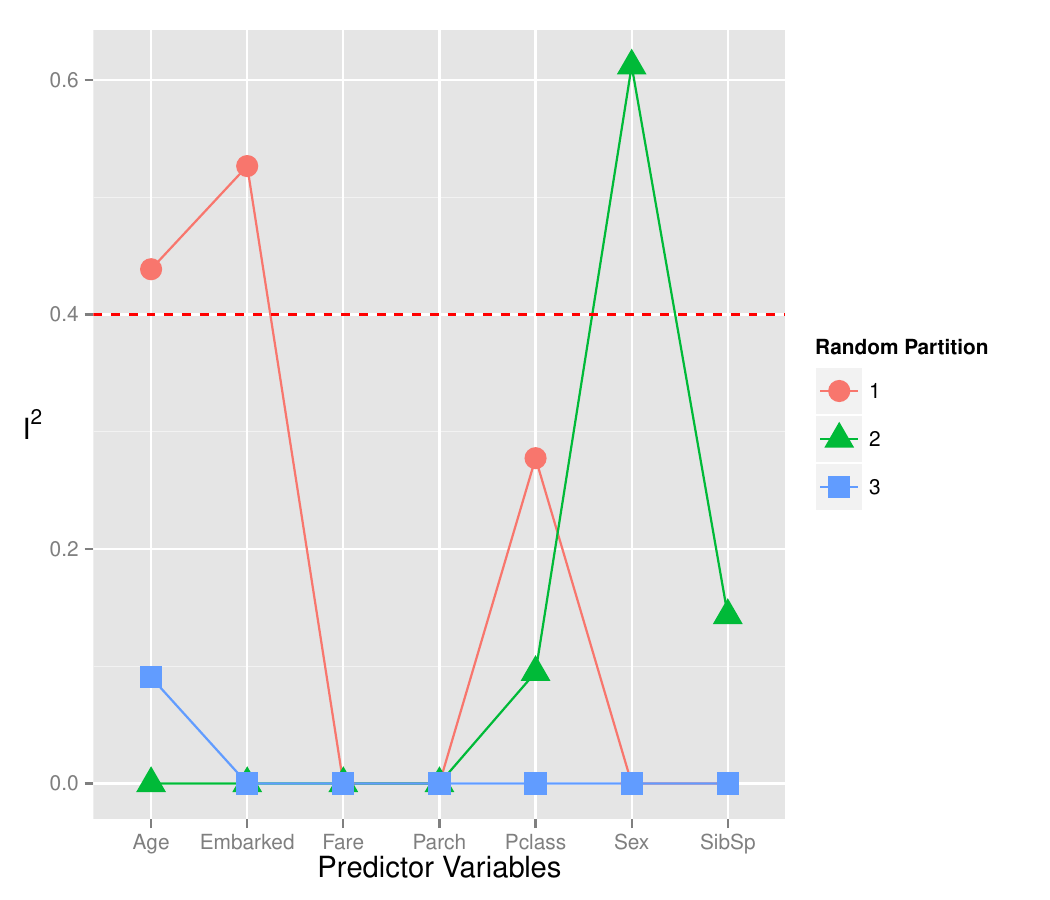}\\
\includegraphics[height=2.3in,width=4in]{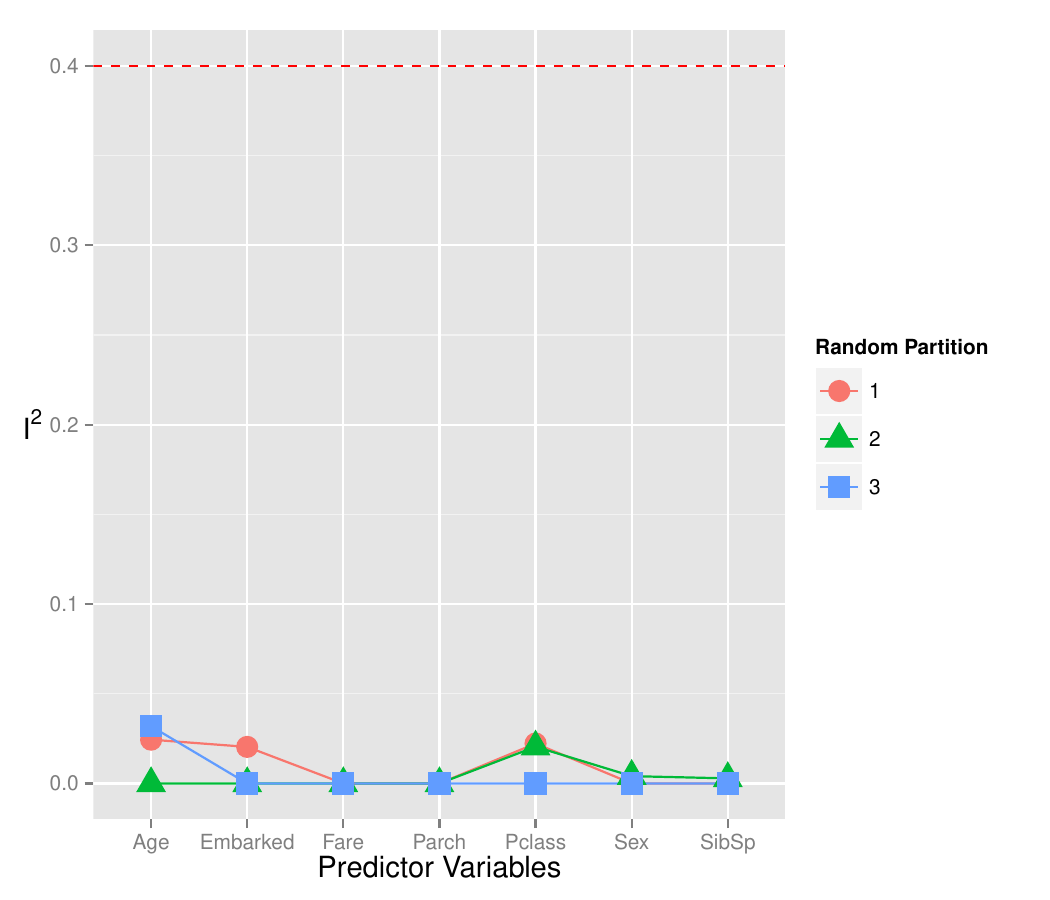}
\caption{(a) Top: $I^2$ diagnostics for three random partitions of the {\it Titanic} dataset (b) Bottom: $I^2$ diagnostic with $\tau^2$ regularization on the {\it Titanic} dataset for three random partitions.}
 \label{fig:tit3}
\end{figure}
 
\begin{figure} [th]
\centering
\includegraphics[height=2in,width=4in]{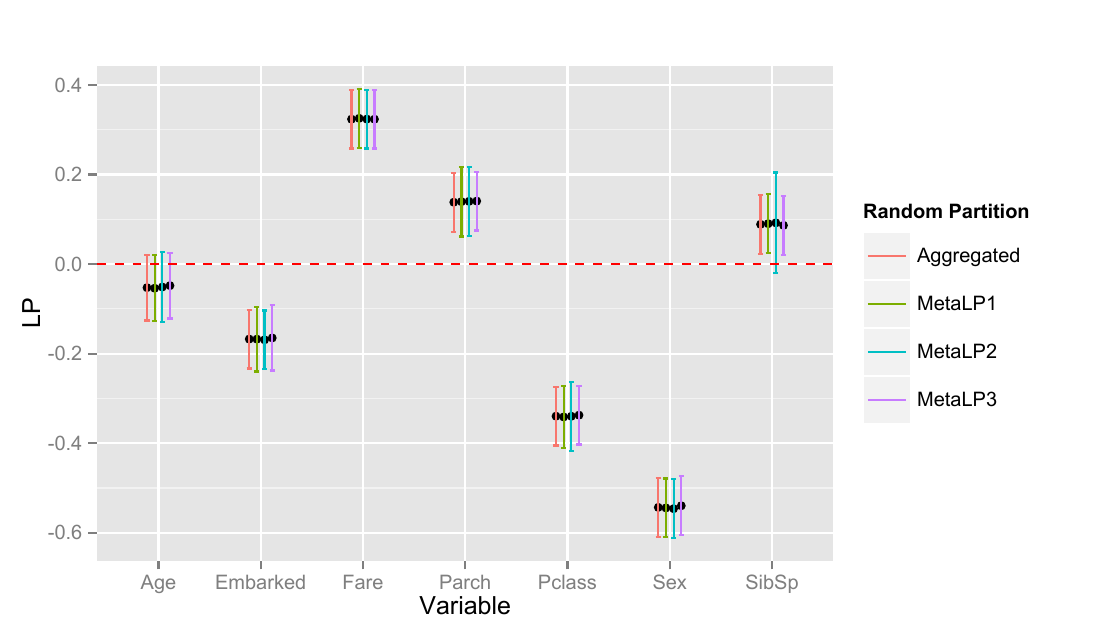}
\caption{95\% Confidence Interval of $\LP$ Statistic for each variable based on three \texttt{MetaLP} repetitions and aggregated full dataset (which is the oracle estimate).}
\label{fig:tit1}
\end{figure}




To develop an automatic solution to the mixed data problem, we start by constructing LP score polynomials for each variable. Figure \ref{fig:SF} shows the shapes of LP basis functions for two variables from the {\it Titanic} data. Next, we randomly assign $891$ observations to $5$ different subpopulations and calculate $\LP$ statistics for each variable in each subpopulation, and then combine $\LP$ statistics to get a combined $\LP$ statistic for each variable. We repeat this process three times to see how much our final \texttt{MetaLP} result changes with different random partitions of the full data. Figures \ref{fig:tit3}(a) shows the $I^2$ statistics for three random partitions on the {\it Titanic} dataset. Even with the randomly assigned partitions, some variables may exhibit heterogeneity among subpopulations as $I^2$ statistics move above 40\%.   For example, random partition 2 results show heterogeneity in variables Embarked and Sex. Thus, we use $\tau^2$ regularization to handle the problem. Figure \ref{fig:tit3}(b) shows the $I^2$ statistics after $\tau^2$ regularization. The additional $\tau^2$ parameter accounts for the heterogeneity in the subpopulations and adjusts the estimators accordingly, resulting in significantly lower $I^2$ statistics for all variables under this model. 

\begin{table*}[ht!]
\centering
\small
\scalebox{1}{
\begin{tabular}{|llll|}
\hline
Variable Name & Type        & Description                       & Values                                                                                        \\ \hline
Survival      & Binary      & Survival                          & 0 = No; 1 = Yes                                                                              \\
Pclass        & Categorical & Passenger Class                   & 1 = 1st; 2 = 2nd; 3 = 3rd                                                                    \\
Sex           & Binary      & Sex                               & Male; Female                                                                                 \\
Age           & Continuous    & Age                               & 0 - 80                                                                                       \\
Sibsp         & Discrete     & Number of Siblings/Spouses Aboard & 0 - 8                                                                                        \\
Parch         & Discrete     & Number of Parents/Children Aboard & 0 - 6                                                                                        \\
Fare          & Continuous     & Passenger Fare                    & 0 - 512.3292                                                                                 \\
Embarked      & Categorical & Port of Embarkation               & \begin{tabular}[c]{@{}l@{}}C = Cherbourg; \\ Q = Queenstown; \\ S = Southampton\end{tabular} \\ \hline
\end{tabular}
} \vskip1em
\caption{Data dictionary for the {\it Titanic} dataset.}
\label{tab:Tit.var}
\end{table*}




Figure \ref{fig:tit1} contains the $\LP$ statistics and their 95\% confidence intervals generated from our algorithm for 3 repetitions of random groupings ($k=5$) along with the confidence intervals generated using the whole dataset.  A \textit{remarkable result of our method} is that the \texttt{MetaLP} estimators and the aggregated (full data) LP estimators are almost indistinguishable for \textit{all} variables. 
In summary, the estimators from our \texttt{MetaLP} method produces very similar inference to the estimators using the entire dataset, which means we can obtain accurate and robust statistical inference while taking advantage of the computational efficiency in parallel, distributed processing.



\section{Simpson's and Stein's Paradox: A MetaLP Perspective}\label{sec:paradox}
Heterogeneity is not solely a big data phenomenon; it can easily arise in the small data setup.  We show two examples, Simpson's Paradox and Stein's Paradox, where blind aggregation \textit{without paying attention to the underlying heterogeneity} leads to a misleading conclusion.

\subsection{Simpson's Paradox}
\label{sec:simpson}
Table \ref{tab:ucb1} shows the UC Berkeley admission data \citesec{Bickel1975} by department and gender.  Looking only at the university level admission rates at the bottom of this table, there appears to be a significant difference in admission rates for males at 45\% and females at 30\%.  However, the department level data \textit{does not} appear to support a strong gender bias as in the university level data.  The real question at hand is whether \emph{there is a gender bias in university admissions?} We provide a concrete statistical solution to the question put forward by \citesec{Pearl2014} regarding the validity and applicability of traditional statistical tools in answering the real puzzle of Simpson's Paradox: ``So in what sense do B-K plots, or ellipsoids, or vectors display, or regressions etc. contribute to the puzzle? They don't. They can't. Why bring them up? Would anyone address the real puzzle? It is a puzzle that cannot be resolved in the language of traditional statistics.''

In particular, we will demonstrate how adopting the \texttt{MetaLP} modeling and combining strategy (that properly takes the existing heterogeneity into account) can resolve issues pertaining to Simpson's paradox \citesec{Simpson1951}. This simple example teaches us that \textit{simply averaging} as a means of combining effect sizes is  \textit{not appropriate} regardless of the size of the data. The calculation for the  weights \textit{must} take into account the underlying departure from homogeneity, which is ensured in the \texttt{MetaLP} distributed inference framework. Now we explain how this  paradoxical reversal can be resolved using the \texttt{MetaLP} approach.

\begin{table}[ht]
\centering
\def\arraystretch{1.5}
\begin{tabular}{l l l}
Dept & Male & Female \\\hline
A & 62\% (512 / 825) & \textbf{82\%} (89 / 108)\\
B &  63\% (353 / 560) & \textbf{68\%} (17 / 25)\\
C & \textbf{37\%} (120 / 325) & 34\% (202 / 593) \\
D & 33\% (138 / 417) & \textbf{35\%} (131 / 375) \\
E & \textbf{28\%} (53 / 191) & 24\% (94 / 393) \\
F & \textbf{6\%} (22 / 373) & 7\% (24 / 341) \\
All & \textbf{45\%} (1198 / 2691) & 30\% (557 / 1835) \\
\hline
\end{tabular} 
\caption{UC Berkeley admission rates by gender by department.}
\label{tab:ucb1}
\end{table}

As both admission ($Y$) and gender ($X$) are binary variables, we can compute at most one LP orthogonal polynomial for each variable $T_1(Y;Y)$ and $T_1(X;X)$; accordingly, we can compute only the first-order linear LP statistics, $\LP[1;Y,X],$ for each department. Following Equation \eqref{eq:cd}, we derive and estimate the aCD for the $\LP$ statistic for each of the $6$ departments, $H(\LP_l[1;X,Y])$, $l=1,\ldots,6$, and for the aggregated university level dataset, $H(\LP_a[1;X,Y])$.  As noted in Section \ref{sec:CD}, the department level aCDs are normally distributed with a mean of $\widehat{\LP}_l[1;X,Y]$ and variance of $1/n_{\ell}$ where $n_{\ell}$ is the number of applicants to department $\ell$.  Similarly, the aggregated aCD is also normally distributed with a mean of $\widehat{\LP}_a[1;X,Y]$ and variance of $1/n_a$ where $n_a$ is the number of applicants across all departments.

Now we apply the heterogeneity-corrected \texttt{MetaLP} algorithm following Theorem \ref{thm:taulp} to estimate the combined aCD across all departments as follows:

\begin{align*} 
&H^{(c)}(\LP[1;X,Y]) = \\
&\Phi \left[ \left( \sum\limits_{\ell=1}^6 \frac{1}{\tau^2+(1/n_{\ell})} \right)^{1/2} (\LP[1;X,Y] - \widehat{\LP}^{(c)}[1;X,Y]) \right] 
\end{align*} 
with
\begin{equation*}
\widehat{\LP}^{(c)}[1;X,Y]) = \frac { \sum_{\ell=1}^{6} (\tau^2+(1/n_{\ell}))^{-1} \widehat{\LP}_\ell[1;X,Y])}
{ \sum_{\ell=1}^{6} (\tau^2+(1/n_{\ell}))^{-1}}
\end{equation*}
where $\widehat{\LP}^{(c)}[1;X,Y])$ and $\sum_{l=1}^{6} (\tau^2+(1/n_{\ell}))^{-1}$ are the mean and variance respectively of the meta-combined aCD for $\LP[1;X,Y]$.  Here, the heterogeneity parameter, $\tau^2$, is estimated using the restricted maximum likelihood formulation outlined in Supplementary Section \ref{sec:tau2estimator}.  Figure \ref{fig:UCB1}(a) displays the estimated aCDs for each department, aggregated data, and for the \texttt{MetaLP} method.  First note that the aggregated data aCD is very different from the department level aCDs, which is characteristic of the Simpson's paradox reversal phenomenon due to naive ``aggregation bias''.  This is why the aggregated data inference suggests a gender bias in admissions, while the department level data does not.  Second, note that the aCD from the \texttt{MetaLP} method provides an estimate that falls more in line with the department level aCDs.  This highlights the advantage of the \texttt{MetaLP} meta-analysis framework for combining information in a judicious manner.  Also, as mentioned in Section \ref{sec:CD}, all traditional forms of statistical inference (e.g. point and interval estimation, hypothesis testing) can be derived from the aCD above.

\begin{figure}[ttt]
\centering
\hspace{.5cm}\includegraphics[height=6cm, width=8.2cm]{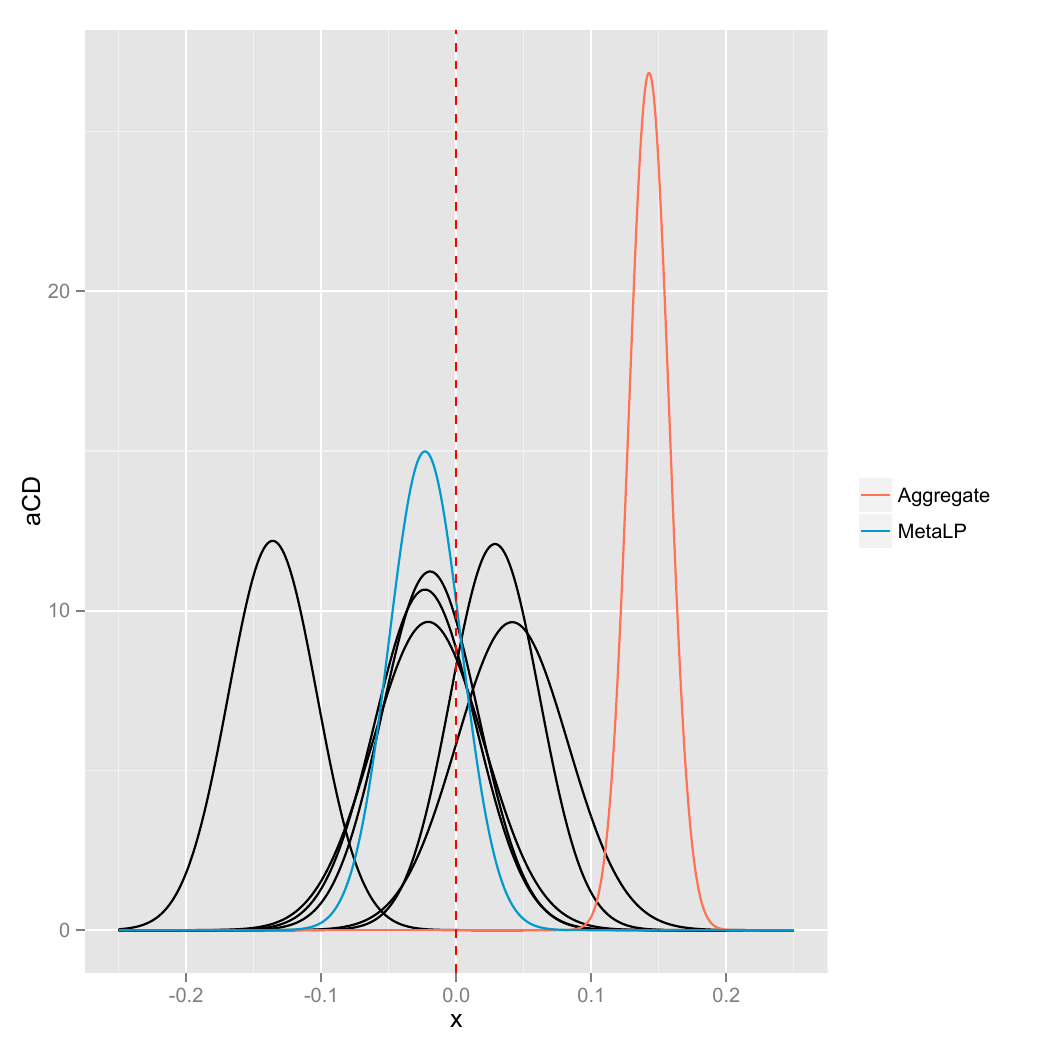}\\
 \hspace{-1.25cm}\includegraphics[height=6cm, width=7cm]{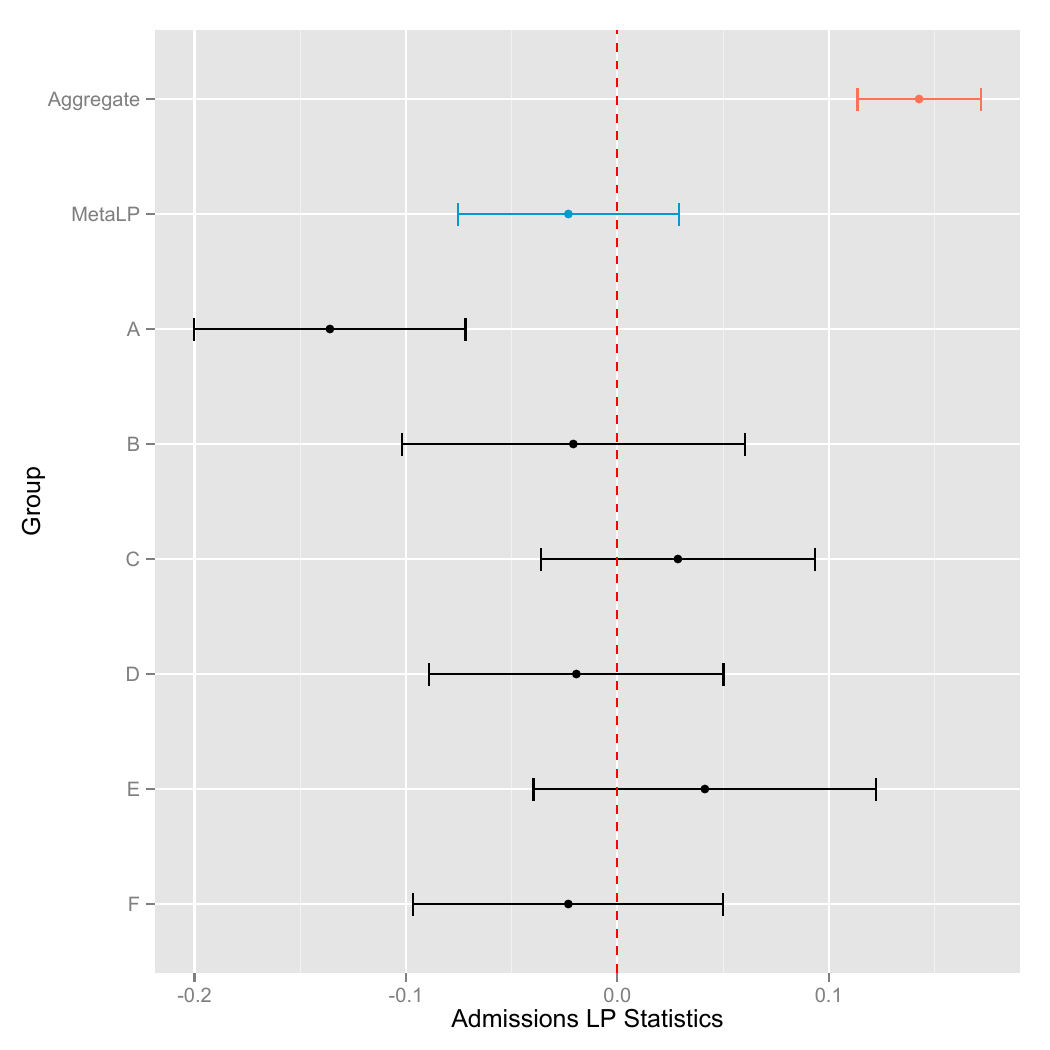}
 \caption{(a) aCDs (top) and (b) 95\% confidence intervals (bottom) for linear $\LP$ statistics for UC Berkeley admission rates by gender (department level aCDs and confidence intervals in black).}
 \label{fig:UCB1}
\end{figure}

For example, we can test $H_0:\LP^{(c)}[1;X,Y]\le 0 $ (indicating no male preference in admissions) vs. $H_1:\LP^{(c)}[1;X,Y] > 0$ (indicating a male preference in admissions) using the aCD for $\LP^{(c)}[1;X,Y]$.  The corresponding p-value for the test comes from the probability associated with the support of $H_0$, $C= (-\infty,0]$, (i.e. ``high'' support value for $H_0$ leads to acceptance) following \citesec{XieSingh2013supp}.  Hence, the p-value for the above test is

\begin{align*}
\text{p-value} &=H\big (  0;  \LP^{(c)}[1;X,Y]  \big) \\
&=\Phi\left(\frac{0-\widehat{\LP}^{(c)}[1;X,Y]}{\sqrt{\sum_{l=1}^{6} (\tau^2+(1/n_l))^{-1}}}\right) \approx .81.
\end{align*}

In this case, the support of the LP CD inference (also known as `belief' in fiducial literature \citesec{KendallStuart1974}) is .81. Hence, at the 5\% level of significance, we fail to reject $H_0$ and confirm that there is no evidence to support a significant gender bias favoring males in admissions using the \texttt{MetaLP} approach.

In addition, we can also compute the 95\% confidence intervals for the $\LP$ statistics measuring the significance of the relationship between gender and admissions as shown in Figure \ref{fig:UCB1}(b).  Note the paradoxical reversal as $5$ out of the $6$ departments show no significant gender bias at the 5\% level of significance (confidence intervals include positive and negative values), while the confidence interval for the aggregated dataset indicates a significantly higher admission rate for males.  On the other hand, note that the \texttt{MetaLP} approach resolves the paradox (which arises \textit{due to the failure of recognizing} the presence of heterogeneity among department admission patterns)  and correctly concludes that no significant gender bias exists (as the confidence interval for the \texttt{MetaLP}-based $\LP$ statistic includes the null value $0$).  

\subsection{Stein's Paradox}
\label{sec:stein}
Perhaps the most popular and classical dataset for Stein's paradox is given in Table~\ref{tab:stein1}, which shows the batting averages of 18 major league players through their first $45$ official at-bats of the $1970$ season. The goal is to predict each player's batting average over the remainder of the season (comprising about $370$ more at bats each) using only the data of the first $45$ at-bats. Stein's shrinkage estimator \citesec{james1961}, which can be interpreted as an empirical Bayes estimator \citesec{EfronMorris1975} turns out to be more than $3$ times more efficient than the MLE estimator. Here we provide a \texttt{MetaLP} approach to this problem by recognizing the ``parallel" structure (18 parallel sub-populations) of baseball data, which fits nicely into the ``decentralized" \texttt{MetaLP} modeling framework.

\begin{table}[th]
\centering
\def\arraystretch{1.2}
\begin{tabular}{l l l l l l}
Name & hits$/$AB & $\hat{\mu}_i^{(\MLE)}$ & $\mu_i$ & $\hat{\mu}_i^{(JS)}$ & $\hat{\mu}_i^{(\LP)}$\\\hline
Clemente & $18/45$ & .400 & .346 & \textbf{.294} & .276 \\
F Robinson & $17/45$ & .378 & .298 & \textbf{.289} & .274 \\
F Howard & $16/45$ & .356 & .276 & .285 & \textbf{.272} \\
Johnstone & $15/45$ & .333 & .222 & .280 & \textbf{.270} \\
Berry & $14/45$ & .311 & .273 & \textbf{.275} & .268 \\
Spencer & $14/45$ & .311 & .270 & .275 & \textbf{.268} \\
Kessinger & $13/45$ & .289 & .263 & .270 & \textbf{.265} \\
L Alvarado & $12/45$ & .267 & .210 & .266 & \textbf{.263} \\
Santo & $11/45$ & .244 & .269 & \textbf{.261} & \textbf{.261} \\
Swoboda & $11/45$ & .244 & .230 & \textbf{.261} & \textbf{.261} \\
Unser & $10/45$ & .222 & .264 & .256 & \textbf{.258} \\
Williams & $10/45$ & .222 & .256 & \textbf{.256} & .258 \\
Scott & $10/45$ & .222 & .303 & .256 & \textbf{.258} \\
Petrocelli & $10/45$ & .222 & .264 & .256 & \textbf{.258} \\
E Rodriguez & $10/45$ & .222 & .226 & \textbf{.256} & .258 \\
Campaneris & $9/45$ & .200 & .286 & .252 & \textbf{.256} \\
Munson & $8/45$ & .178 & .316 & .247 & \textbf{.253} \\
Alvis & $7/45$ & .156 & .200 & \textbf{.242} & .251 \\
\hline
\end{tabular} 
\caption{Batting averages $\hat{\mu}_i^{(\MLE)}$ for 18 major league players early in the 1970 season; $\mu_i$ values are averages over the remainder of the season.  The James-Stein estimates $\hat{\mu}_i^{(JS)}$ and MetaLP estimates $\hat{\mu}_i^{(\LP)}$ provide much more accurate overall predictions for the $\mu_i$ values compared to MLE. MSE ratio for $\hat{\mu}_i^{(JS)}$ to $\hat{\mu}_i^{(\MLE)}$ is $0.283$ and MSE ratio for $\hat{\mu}_i^{(\LP)}$ to $\hat{\mu}_i^{(\MLE)}$ is $0.293$ showing comparable efficiency.
}
\label{tab:stein1}
\end{table}

\begin{figure}[ht]
 \centering
 \includegraphics [scale=.5]{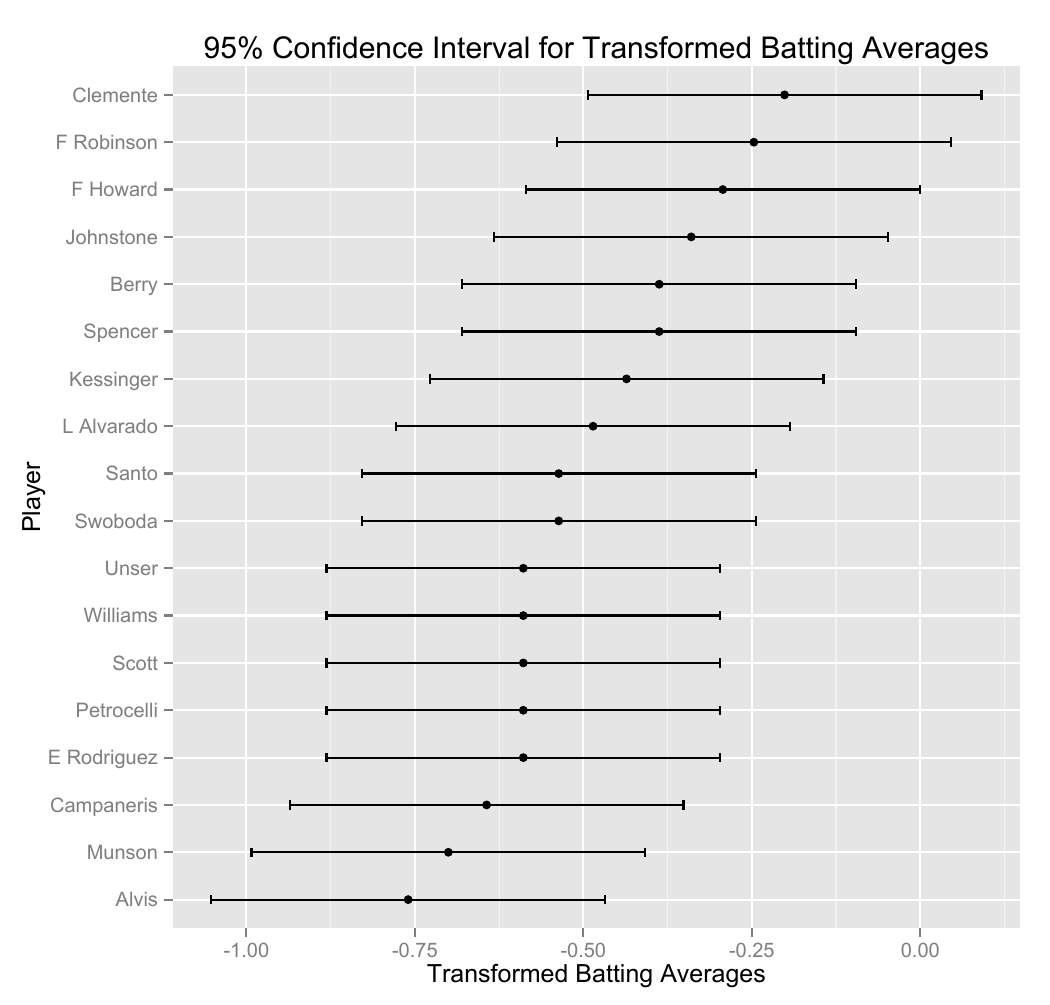}
  \caption{95\% confidence intervals for transformed batting averages, $\theta_i$, for each player, indicating the heterogeneity of the effect sizes estimates.}
  \label {fig:batavg}
\end{figure}

We start by defining the variance-stabilized effect-size estimates for each group

$$\widehat \theta_i = \sin^{-1}(2\hat{\mu}_i^{(\MLE)}-1),~~ \,i=1,\ldots,k ~~~$$

\noindent whose asymptotic distribution is normal with mean $\theta_i$ and variance $1/n_i$ where $n_i=45$ (for all $i$) is the number of at-bats for each player and $\hat{\mu}_i^{(\MLE)}$ is the individual batting average for player $i$. Figure \ref{fig:batavg} provides some visual evidence of the heterogeneity between the studies. 

We apply a \texttt{MetaLP} procedure that incorporates inter-study variations and is applicable for unequal variance/sample size scenarios with no further adjustment.  First, we estimate the weighted mean, $\hat{\theta}_{\mu}$, of the transformed batting averages with weights for each study $(\hat\tau_{{\rm DL}}^2 + n_i^{-1})^{-1}$, where $\hat\tau_{{\rm DL}}^2$ denotes the DerSimonian and Laird data-driven estimate given in Supplementary Section \ref{sec:tau2estimator}.  The \texttt{MetaLP} estimators, $\hat{\theta}_i^{(\LP)}$, are represented as weighted averages between the transformed batting averages and $\hat{\theta}_{\mu}$ as follows:

$$\hat{\theta}_i^{(\LP)} = \lambda\hat{\theta}_{\mu} + (1-\lambda)\widehat \theta_i,~~ ~~(i=1,\ldots,18),~~~~$$

\noindent where $\lambda = (n_i^{-1})/(\hat\tau_{{\rm DL}}^2 + n_i^{-1})$. Table~\ref{tab:stein1} shows that \texttt{MetaLP}-based estimators are as good as James-Stein empirical Bayes estimators for the baseball data. This stems from the simple fact that random-effect meta-analysis and the Stein formulation are mathematically equivalent. But nevertheless, the framework of understanding and interpretations are different. Additionally, \texttt{MetaLP} is much more flexible and automatic in the sense that it works for `any' estimators (such as mean, regression function, classification probability) beyond mean and Gaussianity assumptions. We feel the \texttt{MetaLP} viewpoint is also less mysterious and clearly highlights the core issue of heterogeneity. Our analysis indicates an exciting frontier of future research at the interface of \texttt{MetaLP}, Empirical Bayes, and Stein's Paradox to develop new theory of distributed massive data modeling.

\section{MapReduce Computation Framework and R Functions}
\label{sec:mapreducecode}
In this note, we describe how the proposed \texttt{MetaLP} statistical algorithmic framework for big data analysis can easily be integrated with the \texttt{MapReduce} computational framework, along with the required \texttt{R} code. MapReduce implementation of \texttt{MetaLP} allows efficient parallel processing of large amounts of data to achieve scalability.
\subsection{LP.Mapper}
We apply the following \verb"LP.Mapper" function to each subpopulation. This function computes $\LP[j;X,Y]$ for $j=1,\ldots,m$ (where user selects $m$, which should be less than the number of distinct values of the given random sample). The first step is to design the data-adaptive orthogonal LP polynomial transformation of the given random variable $X$.  This is implemented using the function \verb"LP.Score.fun". The second step uses the LP inner product to calculate the LP variable selection statistic using the function \verb"LP.VarStat" (see Section \ref{sec:LP} for details).

\textbf{Inputs} of \verb"LP.Mapper". $Y$ is binary (or discrete multinomial) and $X$ is a mixed (discrete or continuous type) predictor variable.

\textbf{Outputs} of \verb"LP.Mapper". It returns the estimated $\widehat{\LP}[j;X,Y]$ and the corresponding (asymptotic) sample variance.  Note that the sample LP statistic converges to $\mathcal{N}(0,\sigma_\ell^2=1/n_\ell)$, where $n_\ell$ is the effective sample size of the $\ell$th subpopulation. By effective size we mean $n_\ell - M_\ell(X)$, where $M_\ell(X)$ denotes the number of missing observations for variable $X$ in the $\ell$th partition.  \verb"LP.Mapper" returns only $\{\widehat{\LP}[1;X,Y],\ldots, \widehat{\LP}[m;X,Y]\}$ and $n_\ell$, from which we can easily reconstruct the CD of the LP statistics.

\begin{scriptsize}
\begin{verbatim}

LP.Mapper <- function (Y,x,m=1) {
	LP.Score.fun <- function(x,m){
    	u <- (rank(x,ties.method = c("average"))-.5)/length(x); 
    	m <- min(length(unique(u ))-1, m);
    	S.mat <- as.matrix(poly(u,df=m)); 
    	return(as.matrix(scale(S.mat)))
    }
  	LP.VarStat <- function(Y,x,m){ 
    	x <- ifelse (x=="NULL",NA,x);
    	x <- na.omit (x);
    	if (length (unique(x)) <=1 ){ 
      		r.lp=0;
            n=0;
    	}else{
      		which <- na.action(x);
      		if (length(which)>0) Y <- Y[-which];
      		if (length(unique(Y))<=1){
        		r.lp=0;
                n=0;
      		}else{
        		x <- as.numeric (x);
        		S <- LP.Score.fun(x,m);
        		r.lp <- cor(Y,S);n=length(Y);
      		}
    	}
    	return(c(r.lp,n))
	}
    temp <- LP.VarStat(Y,x,m); 
    output.LP <- temp[1:length(temp)-1];
    output.n <- temp [length(temp)]
    logic <- ifelse (length(temp)-1==m,"NA",
    	"m is not less than the number of distinct value of x")
	return (list(LP=output.LP,n=output.n,Warning=logic)) 
    }
\end{verbatim}
\end{scriptsize}

\subsection{Meta.Reducer}
\verb"LP.Mapper" computes the sample LP statistics and the corresponding sample variance. Now at the `Reduce' step, our goal is to judiciously combine these estimates from $k$ subpopulations to produce the statistical inference for the original large data. Here we implement the MetaReduce strategy to combine the inference from all the subpopulations, implemented in the function  \verb"Meta.Reducer".

Before performing the \texttt{Meta.Reducer} step, we run the `combiner' operation that gathers the outputs of the \texttt{LP.Mapper} function for all the subpopulations and organizes them in the form of a list, which has two components: (i) a matrix \verb"L.value" of order $k \times p$, where $k$ is the number of subpopulations and $p$ is the number of predictor variables (the $(\ell, i)$th element of that matrix stores the $j$th LP statistic $\LP[j;X_i,Y]$ for $\ell$th partition); (ii) a matrix \verb"P.size" of size $k \times p$ ($(\ell, i)$th element stores the effective size of the subpopulation for the variable $\ell$).

\textbf{Inputs} of \verb"Meta.Reducer"
\begin{enumerate}
	\item \verb"L.value" and \verb"P.size"
    \item \verb"fix": a binary argument (TRUE or FALSE), indicating whether to ignore the $\tau^2$ regularization. If it equals to FALSE, then the model with $\tau^2$ regularization is applied.
	\item \verb"method": It's valid only if \verb"fix" equals FALSE, and can equal to either \texttt{"DL"} or \texttt{"REML"}, indicating the estimation method of $\tau^2$.
	\item \texttt{"DL"} stands for the method proposed in \citesec{DerSimonianLaird1986supp}, and \texttt{"REML"} is the restricted maximum likelihood method, which was proposed in \citesec{Normand1999}. We include the calculation methods of these two $\tau^2$ estimators in the next section.
\end{enumerate}


\textbf{Outputs} of \verb"Meta.Reducer"
\begin{enumerate}
  \item Meta-analysis combined LP statistic estimators
  \item Standard errors of meta-analysis combined LP statistic estimators
  \item $I^2$ heterogeneity diagnostic 
  \item $\tau^2$ estimate only if \verb"fix" equals to FALSE
\end{enumerate}

\begin{scriptsize}
\begin{verbatim}
Meta.Reducer <- function(L.value, P.size, fix, method){
	th_c <- NA;
    sd_th_c <- NA;
  	for (i in 1:ncol(L.value)){
    	th_c[i] <- sum(L.value[,i]*P.size[,i])/sum(P.size[,i]);
    	sd_th_c[i] <- sqrt(1/sum(P.size[,i]));
  	}
  	Q <- matrix (,ncol(L.value),1);
  	for (i in 1:ncol(L.value)){
    	Q[i,] <- sum ( P.size[,i]*(L.value [,i] - th_c[i])^2);
  	}
  	K<-NA;
  	for (i in 1:ncol(L.value)){
    	A <- P.size[,i]; 
    	K[i] <- length (A[A!=0]);
  	} 
  	if (fix==T){  
    	I_sq.f <- ifelse ((Q-(K-1))/Q>0, (Q-(K-1))/Q,0);
    	return (list(LP.c=th_c, SE.LP.c=sd_th_c,I_sq.f=I_sq.f))
  	}else{
    if (method=="DL"){
    	tau.sq <- NA;
      for (i in 1:ncol(L.value)){
      	tau.sq[i] <- (Q[i]-(K[i]-1)) / 
        (sum(P.size[,i]) 
        - sum((P.size[,i])^2)/sum(P.size[,i]));
      }
      tau.sq <- ifelse(tau.sq>0,tau.sq,0);
      w_i <- matrix(NA,nrow(P.size), ncol(P.size));
      for (i in 1:ncol(L.value)){
        w_i[,i] <- (1/P.size[,i]+tau.sq[i])^-1;
      }
      mu.hat <- NA;
      SE_mu.hat <- NA;
      for (i in 1:ncol(L.value)){
        mu.hat[i] <- sum(L.value[,i]*w_i[,i])/sum(w_i[,i]);
        SE_mu.hat[i] <- sqrt(1/sum(w_i[,i]));
      }
      lam_i <- matrix (NA,nrow(P.size),ncol(P.size));
      for (i in 1:ncol(L.value)){
        lam_i[,i] <- (1/P.size[,i])/(1/P.size[,i]+tau.sq[i]);
      }
      th.tilde <- matrix(NA,nrow(L.value), ncol(L.value))
      for (i in 1:ncol(L.value)){
        th.tilde[,i] <- lam_i[,i] * mu.hat [i] + 
        (1-lam_i[,i])*L.value[,i];
      }
      th.tilde <- ifelse(is.nan(th.tilde)==T,0,th.tilde);
      Q <- matrix (NA,ncol(L.value),1);
      for (i in 1:ncol(L.value)){
        Q[i,] <- sum ( w_i[,i]*(th.tilde [,i] - mu.hat[i])^2);
      }
      I_sq.r <- ifelse ((Q-(K-1))/Q>0, (Q-(K-1))/Q,0);
      return (list (LP.c=mu.hat,
      SE.LP.c=SE_mu.hat,I_sq.r=I_sq.r,tau.sq=tau.sq))
    }
    if (method=="REML"){
      tau.sq <- NA;
      for (i in 1:ncol(L.value)){
        tau.sq[i] <- (Q[i]-(K[i]-1)) / 
        (sum(P.size[,i]) - 
        sum((P.size[,i])^2)/sum(P.size[,i]))
      }
      tau.sq <- ifelse(tau.sq>0,tau.sq,0);
      for (i in 1:ncol(L.value)){
      	if (sum(P.size[,i]==0)>0){
        	n <- P.size[,i][-which(P.size[,i]==0)];
        	thh <- L.value[,i][-which(P.size[,i]==0)];
        }else{
        	n <- P.size[,i];
        	thh <- L.value[,i]; 
        }
        nloop <- 0;
        absch <- 1;
        while (absch > 10^(-10)){
        	nloop <- nloop + 1;
        	if (nloop > 10^5){
        		tau.sq[i] <- NA ; 
          	}
          	else{
            tau.sq.old <- tau.sq[i] 
            # update thetaR, wR
            wR <- 1/(1/n + tau.sq.old);
            thetaR <- sum(wR*thh) / sum(wR);
            # update tauR
            tau.sq[i] <- sum(wR^2*(K[i]/(K[i]-1)*
            (thh- thetaR)^2 - 1/n) ) / sum(wR^2);
            absch <- abs(tau.sq[i] - tau.sq.old);
          	}
        }
      }
      tau.sq <- ifelse(tau.sq>0, tau.sq, 0);
      w_i <- matrix(NA,nrow(P.size),ncol(P.size));
      for (i in 1:ncol(L.value)){
      	w_i[,i] <- (1/P.size[,i]+tau.sq[i])^-1;
      }
      mu.hat <- NA;
      SE_mu.hat <- NA;
      for (i in 1:ncol(L.value)){
        mu.hat[i] <- sum(L.value[,i]*w_i[,i])/sum(w_i[,i]);
        SE_mu.hat[i] <- sqrt(1/sum(w_i[,i]));
      }
      lam_i <- matrix(NA,nrow(P.size), ncol(P.size));
      for (i in 1:ncol(L.value)){
        lam_i[,i] <- (1/P.size[,i])/(1/P.size[,i]+tau.sq[i]);
      }
      th.tilde <- matrix (NA,nrow(L.value),ncol(L.value));
      for (i in 1:ncol(L.value)){
        th.tilde [,i] <- lam_i[,i] * mu.hat [i] + 
        (1-lam_i[,i])*L.value[,i];
      }
      th.tilde <- ifelse(is.nan(th.tilde)==T,0,th.tilde);
      Q <- matrix(NA,ncol(L.value),1);
      for (i in 1:ncol(L.value)){
        Q[i,] <- sum(w_i[,i]*(th.tilde [,i] - mu.hat[i])^2);
      }
      I_sq.r <- ifelse ((Q-(K-1))/Q>0,(Q-(K-1))/Q,0);
      return(list(LP.c=mu.hat,SE.LP.c=SE_mu.hat,
      I_sq.r=I_sq.r,tau.sq=tau.sq))
      }
  	}
} 
\end{verbatim}
\end{scriptsize}
\section{$\tau^2$ Estimator}
\label{sec:tau2estimator}
There are many different proposed estimators for the $\tau^2$ parameter.  We consider the DerSimonion and Laird estimator \citesec{DerSimonianLaird1986supp}, $\hat{\tau}^2_{{\rm DL}}$, and the restricted maximum likelihood estimator \citesec{Normand1999}, $\hat{\tau}^2_{{\rm REML}}$, for our analysis. $\hat{\tau}^2_{{\rm DL}}$ can be found from the following equation: 
\begin{multline*}
\label{eq:dl1}
\hat{\tau}_{DL}^2 = \max \left\{
0,
\frac{Q-(k-1)}
{\sum_\ell  s_\ell ^{-2} - \sum_\ell  s_\ell ^{-4} / \sum_\ell  s_\ell ^{-2}}
\right\};
\end{multline*}
where 
\[ Q= \sum_{\ell =1}^{k} \left(\widehat{\LP}_{\ell }[j;X,Y]-\widehat{\LP}^{(c)}[j;X,Y] \right)^2 s_\ell ^{-2}. \]

However, $\hat{\tau}^2_{{\rm REML}}$ should be calculated in an iterative fashion to maximize the restricted likelihood following these steps:\\\\
\noindent \textbf{Step 1:} Obtain the initial value, $\hat{\tau}_{0}^2$. We use $\hat{\tau}^2_{{\rm DL}}$ as the initial value:\\
\begin{equation*}
\label{eq:r4}
\hat{\tau}_{0}^2 = \hat{\tau}^2_{{\rm DL}}.
\end{equation*}
\textbf{Step 2:} Obtain $\widehat{\LP}_{\tau}^{(c)}[j;X,Y]$ ($\tau$-corrected combined LP statistics).\\

\begin{multline*}
\widehat{\LP}_{\tau}^{(c)}[j;X,Y] =
\frac {\sum_\ell w_\ell(\tau^2_{0})
\widehat{\LP}_{\ell }[j;X,Y]}
{\sum_\ell  w_\ell (\hat{\tau}^2_{0})}; \\
~ w_\ell (\hat{\tau}^2_{0}) = (s_\ell ^2+\hat{\tau}^2_{0})^{-1}.
\end{multline*}
\textbf{Step 3:} Obtain the REML estimate.
\begin{multline*}
\label{eq:r6}
\hat{\tau}_{{\rm REML}}^2 = \\
\frac {\sum_\ell w_\ell^2(\hat{\tau}^2_{0})  \left( \frac{k} {k-1} \left(\widehat{\LP}_{\ell}[j;X,Y] - \widehat{\LP}_{\tau}^{(c)}[j;X,Y] \right) - s_\ell^2 \right)} {\sum_\ell w_\ell^2(\hat{\tau}^2_{0})}.
\end{multline*}
\textbf{Step 4:}  Compute new $\widehat{\LP}_{\tau}^{(c)}[j;X,Y]$ by plugging $\hat{\tau}^2_{{\rm REML}}$ obtained in Step 3 into formula from Step 2.\\\\
\textbf{Step 5:} Repeat Step 2 and Step 3 until $\hat{\tau}^2_{{\rm REML}}$ converges.\\

Convergence can be measured as the absolute difference between
$\hat{\tau}^2_{{\rm REML}}$ from the latest iteration and the previous
iteration reaching a threshold close to zero.

\bibliographystylesec{IEEEtran}
\bibliographysec{IEEEabrv,mybibsupp}

\end{document}